\documentclass[12pt]{article}
\pdfoutput=1

\usepackage{amsmath}
\usepackage{amssymb}
\usepackage{epsfig}
\usepackage{graphicx}
\usepackage{cite}
\usepackage{multirow}
\usepackage{longtable}
\usepackage{lscape}
\usepackage{bbm}
\usepackage{dcolumn}
\usepackage{rotating}
\usepackage[squaren]{SIunits}

\allowdisplaybreaks[4] % for page break

\newcommand{\capdef}{}
\newcommand{\mycaption}[2][\capdef]{\renewcommand{\capdef}{#2}%
        \caption[#1]{{\footnotesize #2}}}
\makeatletter
\renewcommand{\fnum@table}{\textbf{\tablename~\thetable}}
\renewcommand{\fnum@figure}{\textbf{\figurename~\thefigure}}
\makeatother

\newcounter{myenumi}

\renewcommand{\themyenumi}{\roman{myenumi}}
{\end{list}}

\newlength{\myem}
\newcounter{mysubequation}[equation]

\makeatletter
\renewcommand{\section}{\@startsection{section}{1}{0em}{-\baselineskip}%
{\baselineskip}{\normalfont\large\bfseries}}
\renewcommand{\subsection}%
{\@startsection{subsection}{2}{0em}{-0.7\baselineskip}%
{0.7\baselineskip}{\normalfont\bfseries}}
\makeatother

\newcommand{\bi}{\begin{itemize}}
\newcommand{\ei}{\end{itemize}}

\newcommand{\be}{\begin{equation}}
\newcommand{\ee}{\end{equation}}
\newcommand{\bea}{\begin{eqnarray}}
\newcommand{\eea}{\end{eqnarray}}

\newcommand{\ie}{{\it i.e.}}

\newcommand{\eg}{{\it e.g.}}

\newcommand{\cf}{{\it cf.}}

\newcommand{\etc}{{\it etc.}}
\newcommand{\eq}{Eq.}
\newcommand{\eqs}{Eqs.}

\newcommand{\fig}{Fig.}

\newcommand{\Ref}{Ref.}
\newcommand{\Refs}{Refs.}
\newcommand{\Sec}{Sec.}
\newcommand{\Secs}{Secs.}
\newcommand{\App}{App.}

\newcommand{\Tab}{Table}

\newcommand{\equ}[1]{\eq~(\ref{equ:#1})}
\newcommand{\figu}[1]{\fig~\ref{fig:#1}}

\newcommand{\rhat}{{\widehat R}}

\begin{document}
%%%%%%%%%%%%%%%%%%%%%%%%%%%%%%%%%%%%%%%%%%%%%%%%%%%%%%%%%%%%%%%%%%%%%
%%%%                     Title-page                              %%%%
%%%%%%%%%%%%%%%%%%%%%%%%%%%%%%%%%%%%%%%%%%%%%%%%%%%%%%%%%%%%%%%%%%%%%

\begin{titlepage}

\renewcommand{\thefootnote}{\alph{footnote}}

% \vspace*{-3.cm}
% \begin{flushright}
% 
% \end{flushright}

%\vspace*{0.5cm}

\renewcommand{\thefootnote}{\fnsymbol{footnote}}
\setcounter{footnote}{-1}

{\begin{center}
{\large\bf
Systematics in the Interpretation of Aggregated Neutrino Flux Limits and Flavor Ratios from Gamma-Ray Bursts 
} 

\end{center}}

\renewcommand{\thefootnote}{\alph{footnote}}

\vspace*{.8cm}
\vspace*{.3cm}
{\begin{center} {\large{\sc 
                Philipp~Baerwald\footnote[1]{\makebox[1.cm]{Email:}
                philipp.baerwald@physik.uni-wuerzburg.de}, 
                Svenja~H{\"u}mmer\footnote[2]{\makebox[1.cm]{Email:}
                svenja.huemmer@physik.uni-wuerzburg.de}, and
                Walter~Winter\footnote[3]{\makebox[1.cm]{Email:}
                winter@physik.uni-wuerzburg.de}
                }}
\end{center}}
\vspace*{0cm}
{\it
\begin{center}

\footnotemark[1]${}^,$\footnotemark[2]${}^,$\footnotemark[3]
       Institut f{\"u}r Theoretische Physik und Astrophysik, \\ Universit{\"a}t W{\"u}rzburg, 
       97074 W{\"u}rzburg, Germany

\end{center}}

\vspace*{1.5cm}

\begin{center}
{\Large \today}
\end{center}

{\Large \bf
\begin{center} Abstract \end{center}  }

Gamma-ray burst analyses at neutrino telescopes are typically based on diffuse or stacked (\ie, aggregated)  neutrino fluxes, because the number of events expected from a single burst is small. The interpretation of aggregated flux limits implies new systematics not present for a single burst, such as by the integration over parameter distributions (diffuse fluxes), or by the low statistics in small burst samples (stacked fluxes).
We simulate parameter distributions with a Monte Carlo method computing the spectra burst by burst,
as compared to a conventional Monte Carlo integration. With this approach, we can predict
the behavior of the flux in the diffuse limit as well as in low statistics stacking samples, such as used in recent IceCube data analyses. We also include the flavor composition at the detector (ratio between muon tracks and cascades) into our  considerations.
We demonstrate that the spectral features, such as a characteristic multi-peak structure coming from photohadronic interactions, flavor mixing, and magnetic field effects, are typically present even in diffuse neutrino fluxes if only the redshift distribution of the sources is considered, with $z \simeq 1$ dominating the neutrino flux. On the other hand, we show that variations of the Lorentz boost can only be interpreted in a model-dependent way, and can be used as a model discriminator. For example, we illustrate that the observation of spectral features in aggregated fluxes will disfavor the commonly used assumption that bursts with small Lorentz factors dominate the neutrino flux, whereas it will be consistent with the hypothesis that the bursts have similar properties in the comoving frame.
%
% Finally, we analyze the cumulative distribution function of GRBs. For the redshift distribution, the systematical error on the quasi-diffuse flux introduced by a stacking of order 100 bursts, as in recent IceCube analyses, is inferred to be about 50\% (at the 90\% CL).

\vspace*{.5cm}

\end{titlepage}

\newpage

\renewcommand{\thefootnote}{\arabic{footnote}}
\setcounter{footnote}{0}

\section{Introduction}

Neutrino telescopes, such as IceCube~\cite{Ahrens:2003ix} or ANTARES~\cite{Aslanides:1999vq}, are expected to reveal the nature of the sources of the highest-energetic cosmic rays. If a substantial amount of protons (or heavier nuclei) are accelerated in the sources, charged pion production is expected via $pp$ and $p\gamma$ interactions, and therefore a neutrino flux.  There are numerous candidate sources, see \Ref~\cite{Becker:2007sv} for a review and \Ref~\cite{Rachen:1998fd} for the general theory. We focus on the prompt emission of gamma-ray bursts (GRBs) in this study, where photohadronic ($p\gamma$) interactions are expected to lead to a significant flux of neutrinos~\cite{Waxman:1997ti}.  If GRBs are the sources of the highest-energetic cosmic rays, the expected neutrino flux is just around the corner. In fact, recent IceCube-40 data (referring to the 40 string configuration of IceCube) start to test this paradigm~\cite{Abbasi:2011qc,Ahlers:2011jj}, based on analytical estimates of the neutrino fluxes, see \Refs~\cite{Waxman:1997ti,Guetta:2003wi,Becker:2005ej,Razzaque:2006qa} and \Ref~\cite{Abbasi:2009ig} for the application to IceCube data.
On the other hand, numerical calculations indicate that additional processes in the photohadronic interactions, flavor mixing, and magnetic field effects affect normalization and shape of the expected neutrino fluxes, see \Refs~\cite{Murase:2005hy,Lipari:2007su,Baerwald:2010fk}. In \Ref~\cite{Baerwald:2010fk}, we have made the impact of these effects on the original Waxman-Bahcall (WB) flux shape~\cite{Waxman:1997ti} very explicit. For example,
a characteristic multi-peak structure from high-energy processes in the photohadronic interactions, magnetic field effects, and flavor mixing has been predicted for the same astrophysical assumptions as for the WB flux. 

While \Ref~\cite{Baerwald:2010fk} is essentially based on one set of parameters only, we focus in this study on {\em aggregated} neutrino fluxes. Because of the low statistics expected from a single burst, any state-of-the-art analysis is based on such an aggregated flux, which can either be a diffuse flux, for which individual sources are not resolved, or a stacked flux, for which the neutrino fluxes from individual sources are expected to be correlated with their gamma-ray counterparts. 
The interpretation of aggregated flux limits, of course, implies new systematics not present for a single burst, such as by the integration over parameter distributions (for diffuse fluxes), or the assumptions for unknown burst parameters (stacked fluxes). In the stacking case, the time- and directional-wise correlation can be used to suppress backgrounds, and the fluence of the gamma-rays can even be used to compute the expected shape of the neutrino spectrum~\cite{Abbasi:2011qc}. 
 However,  since for many bursts the parameters needed for such an estimate are unknown, such as the redshift or Lorentz factor, ``standard values'' are often used. It is one of the motivations for this study to critically review these standard values from a theoretical perspective, in order to reduce the systematical errors in the interpretation of future analyses. For instance, depending on the method, a too large standard value for the redshift may easily lead to an overestimated neutrino flux.  In addition, we discuss if the spectral multi-peak features are present in aggregated fluxes. Note that experiments as IceCube have limits with optimal sensitivities in specific energy ranges, see \eg\ discussion in \Refs~\cite{Halzen:2010yj,Winter:2011jr}. 
% The best sensitivity can be obtained if a peak in the flux coincides with the differential limit of the instrument. 
Therefore, it is relevant if these spectral features are present in aggregated fluxes, especially in a diffuse flux.

Predictions for diffuse neutrino fluxes have, for instance, been made in \Refs~\cite{Halzen:1999xc,Murase:2005hy,Gupta:2006jm}. For a diffuse flux, typically at least the redshift distribution is integrated over, which is assumed to follow the star formation rate in some form. Apart from the redshift, interesting parameters which vary from burst to burst are Lorentz factor, luminosity, gamma-ray fluence, and variability timescale. For example, 
in \Ref~\cite{Gupta:2006jm}, relying on conventional fireball phenomenology, a log normal distribution for the Lorentz factor is assumed. 
In the conventional fireball model, the main contribution to the neutrino flux is assumed to come from small Lorentz factors, see \Refs~\cite{Guetta:2000ye,Guetta:2001cd} for discussions on the parameter space constraints. This conclusion is based on the estimate of the size of the interaction region from the variability timescale, which strongly depends on the Lorentz factor. However, in a modified version of the fireball model including with the hypothesis that the bursts have similar properties in the comoving frame (shock rest frame) the dominance of the small Lorentz factors is no longer given, as we will demonstrate. In fact, a recent analysis by Ghirlanda et al.~\cite{Ghirlanda:2011bn}, which appeared during completion of this work, favors this hypothesis from the simple argument that the variations of the bulk Lorentz factor imply the Amati~\cite{Amati:2002ny} and Yonetoku~\cite{Yonetoku} relations, while the bursts are expected to be similar in the comoving frame.
In order to clarify the model-dependence  (\cf, \Sec~\ref{sec:source}), we start with the minimal ingredients (and input parameters) to neutrino production needed to numerically reproduce the analytical estimates in  \Refs~\cite{Waxman:1997ti,Guetta:2003wi,Abbasi:2011qc}. We clearly separate these from model-dependent assumptions (\Sec~\ref{sec:obs}), which has the advantage that one can clearly see where the model-dependent hypotheses enter. The main questions which we pursue in this context are: Assuming theoretical expectations for parameter distributions, from which parameters do the main contributions to the neutrino flux come from? As a consequence, what value should one assign to a burst with unknown redshift or Lorentz factor, such as in a stacking analysis?  What are the assumptions going into that conclusion, in particular, which assumptions hold in a model-independent way? In order to discuss these questions in a coherent and transparent way, we discuss each of the relevant input parameters separately in \Sec~\ref{sec:diffuse}.

In principle, a neutrino telescope cannot only measure the neutrino flux, but also distinguish different flavors by different event topologies~\cite{Beacom:2003nh}. In the simplest case,
 one may use two topologies: tracks (mainly from $\nu_\mu$) and cascades
  (mainly induced by $\nu_e$ and $\nu_\tau$) to construct an observable muon track to cascade flavor ratio as a flavor dependent quantity. Note that this flavor ratio is defined in the spirit of cascade searches from extragalactic neutrino sources, which has been very recently conducted in \Ref~\cite{IcCascade:2011ui} for IceCube-22. From the charged pion $\pi \to \mu \to e$ decay chain, the flavor composition at the source is given by  $\nu_e:\nu_\mu:\nu_\tau = 1:2:0$ (``pion beam source''), and, for GRBs, changes to $0:1:0$ at higher energies~\cite{Kashti:2005qa,Lipari:2007su} (``muon damped source'') because of the synchrotron cooling and decay of
the intermediate muons, pions, and kaons. 
Although one may not expect to observe this effect with high statistics on the timescale of ten years of full IceCube running due to the lower effective area for cascades, some conclusions on the astrophysical source (such as the magnetic field~\cite{Hummer:2010ai}) or the particle physics during neutrino propagation (see, \eg, \Ref~\cite{Pakvasa:2008nx} for a short review) may be obtained at IceCube or future potential upgrades. Since the flux normalization drops out of this flavor ratio, it ought to be relatively robust with respect to astrophysical uncertainties~\cite{Mehta:2011qb}.  In this study, we do not only discuss the muon neutrino flux at the detector, but also the flavor ratio. In particular, we will demonstrate that the aggregated neutrino flavor ratio, though perhaps measurable with much lower statistics, is a more reliable prediction for GRBs than the flux itself; see \Sec~\ref{sec:diffuse}.

Finally, we discuss the systematical error on the quasi-diffuse flux introduced by the stacking statistics, \ie, the number of stacked bursts, in \Sec~\ref{sec:lowstatistics}. Note that this systematical error is important for the interpretation with respect to the paradigm that GRBs are the sources of the highest energetic cosmic rays.
It is also illuminating to show what the main qualitative differences between stacked and diffuse fluxes are, and where we expect the borderline. Throughout this paper, we use a Monte Carlo method to compute the neutrino fluxes. Instead of  Monte Carlo integration over the parameter distributions, we compute the neutrino spectrum burst-by-burst, sampling over the input parameter distributions. This method has the advantage that it converges to conventional (Monte Carlo) integration in the limit of a large number of bursts, such as $\mathcal{O}(10 \, 000)$ bursts expected in the observable universe over a timescale of ten years. As we shall see, this limit then also represents a diffuse flux. On the other hand, small sample sizes can be chosen to simulate the impact of stacking analyses of \eg\ $\mathcal{O}(100)$ bursts, as used in \Ref~\cite{Abbasi:2011qc}.

\section{The source model}
\label{sec:source}

Here, we describe the model we use for neutrino production, see also \Ref~\cite{Baerwald:2010fk}. First, in \Sec~\ref{sec:summaryshort}, we give a short summary of the model and discuss the translation into observables. Details can be found in \App~\ref{sec:details}, where also a reference parameter set reproducing the original WB shape is derived,  which we use in the following sections, and the effects of the photohadronic high-energy processes and secondary cooling are shown quantitatively for this example. We also discuss the limitations of the model in greater detail there. This model description
is kept as independent as possible from any underlying astrophysical gamma-ray burst model. In the next step, in \Sec~\ref{sec:obs}, we discuss the connection with observables and we demonstrate where model-dependent relationships enter. 

\subsection{Short model summary}
\label{sec:summaryshort}

The idea of this model is to describe the analytical approaches in \Refs~\cite{Waxman:1997ti,Guetta:2003wi,Abbasi:2009ig} as closely as possible, which are used for state-of-the-art IceCube stacking analyses, while introducing more realistic descriptions for energy losses and escape required to describe magnetic field and flavor effects. As indicated above, we clearly separate the (minimal) set of ingredients needed for the description of the neutrino spectra (\Sec~\ref{sec:summaryshort}) and the methods to estimate these quantities in a model-dependent way (\Sec~\ref{sec:modeldep}).  The ingredients of our model closely follow \Ref~\cite{Hummer:2010ai}, where the main differences is the origin of the target photons (which come from the synchrotron radiation of co-accelerated electrons/positrons in \Ref~\cite{Hummer:2010ai}). Note that in this work, we do not revise the absolute normalization of the expected neutrino flux, as it may be expected from the gamma-ray or cosmic ray counterpart, but focus on the theoretically anticipated relative contributions from different bursts. Instead we use 
for the expected normalization of the (diffuse) flux~\cite{Waxman:1998yy} (updated in \Ref~\cite{Waxman:2002wp})
\begin{equation}
E_\nu^2 \phi_\alpha = 0.45 \cdot 10^{-8} \, \frac{f_\pi}{0.2} \, \giga\electronvolt \, \centi\meter^{-2} \, \second^{-1} \, \steradian^{-1}
\label{equ:wb}
\end{equation}
per neutrino species ($\nu_e$, $\nu_{\mu}$, or $\bar{\nu}_{\mu}$ from the pion decay chain) before flavor mixing, or, to a first approximation, for any flavor after flavor mixing. This normalization depends on the fraction of the proton energy going into pion production $f_\pi$, for which we assume $f_\pi \simeq 0.2$. We will examine the assumptions going into this frequently used estimate elsewhere~\cite{Hummer:prep}.

 The photon spectrum is assumed to fit the observed spectral energy distributions of GRBs described as a broken power law with a spectral break,  parameterized in the shock rest frame (SRF)/comoving frame by 
\begin{equation}
	N'_{\gamma}(\varepsilon') = C'_{\gamma} \cdot \left\{ \begin{array}{ll} \left( \frac{\varepsilon'}{\varepsilon'_{\gamma,\text{break}}} \right)^{\alpha_{\gamma}} & \varepsilon'_{\gamma,\text{min}} \leq \varepsilon' < \varepsilon'_{\gamma,\text{break}} \\ \left( \frac{\varepsilon'}{\varepsilon'_{\gamma,\text{break}}} \right)^{\beta_{\gamma}} & \varepsilon'_{\gamma,\text{break}} \leq \varepsilon' < \varepsilon'_{\gamma,\text{max}} \\ 0 & \text{else} \end{array} \right. \quad , \label{equ:targetphoton}
\end{equation}
where $C'_{\gamma}$ is a normalization factor in $\giga\electronvolt^{-1} \, \centi\meter^{-3}$. Typical values are $\alpha_{\gamma} \approx -1$ and $\beta_{\gamma} \approx -2$.
The proton spectrum is assumed to be a cut-off power law with a spectral index expected from Fermi shock acceleration and a cut-off energy constrained by the synchrotron losses of the protons:
\begin{equation}
	N'_p(E_p') = C'_p \cdot \left\{ \begin{array}{ll} \left( \frac{E_p'}{\mathrm{GeV}} \right)^{\alpha_p} \cdot \exp \left( - \frac{E_p'^{2}}{E'^{2}_{p,\text{max}}} \right) &  E_p' \; \ge \;  E'_{p,\text{min}}  \\ 0 & \text{else} \end{array} \right. , \label{equ:targetproton}
\end{equation}
where we typically choose $\alpha_p = -2$ and $C'_p$ is the normalization of the spectrum (in units of $\giga\electronvolt^{-1} \, \centi\meter^{-3}$).

  Photohadronic interactions lead to charged pion, kaon, and neutron production. These particles decay further through weak decays into muons and neutrinos. 
Our photohadronic interaction model is based on \Ref~\cite{Hummer:2010vx} (model Sim-B), based on the physics of SOPHIA~\cite{Mucke:1999yb}, including direct production, different resonances, and high energy processes. The helicity dependent muon decays are taken into account as in \Ref~\cite{Lipari:2007su}. We assume that the source is optically thin to neutrons, which means that  neutrons produced by the photohadronic interactions will decay and lead to an additional electron antineutrino flux. 
 For all produced charged secondaries, we consider synchrotron energy losses and escape via decay, which leads to a loss steepening of the spectrum at
\begin{equation}
E'_c = \sqrt{ \frac{9 \pi \epsilon_0 m^5 c^7}{\tau_0 e^4 B'^2}} \, ,
\label{equ:ec}
\end{equation}
depending on the properties of the secondaries (mass $m$ and rest frame lifetime $\tau_0$) and the value of $B'$.
These spectral breaks, together with high energy processes in photohadronic interactions, translate into a characteristic multi-peak structure in the neutrino spectra, see \App~\ref{sec:typ}.
Although not all features of the sources may be described by our method (see, \eg, \Ref~\cite{Murase:2005hy}), it has the advantage that the results can be directly related to \Refs~\cite{Waxman:1997ti,Guetta:2003wi,Abbasi:2009ig}.

\begin{figure}[t]
 \includegraphics[width=1\textwidth,viewport=20 192 696 548]{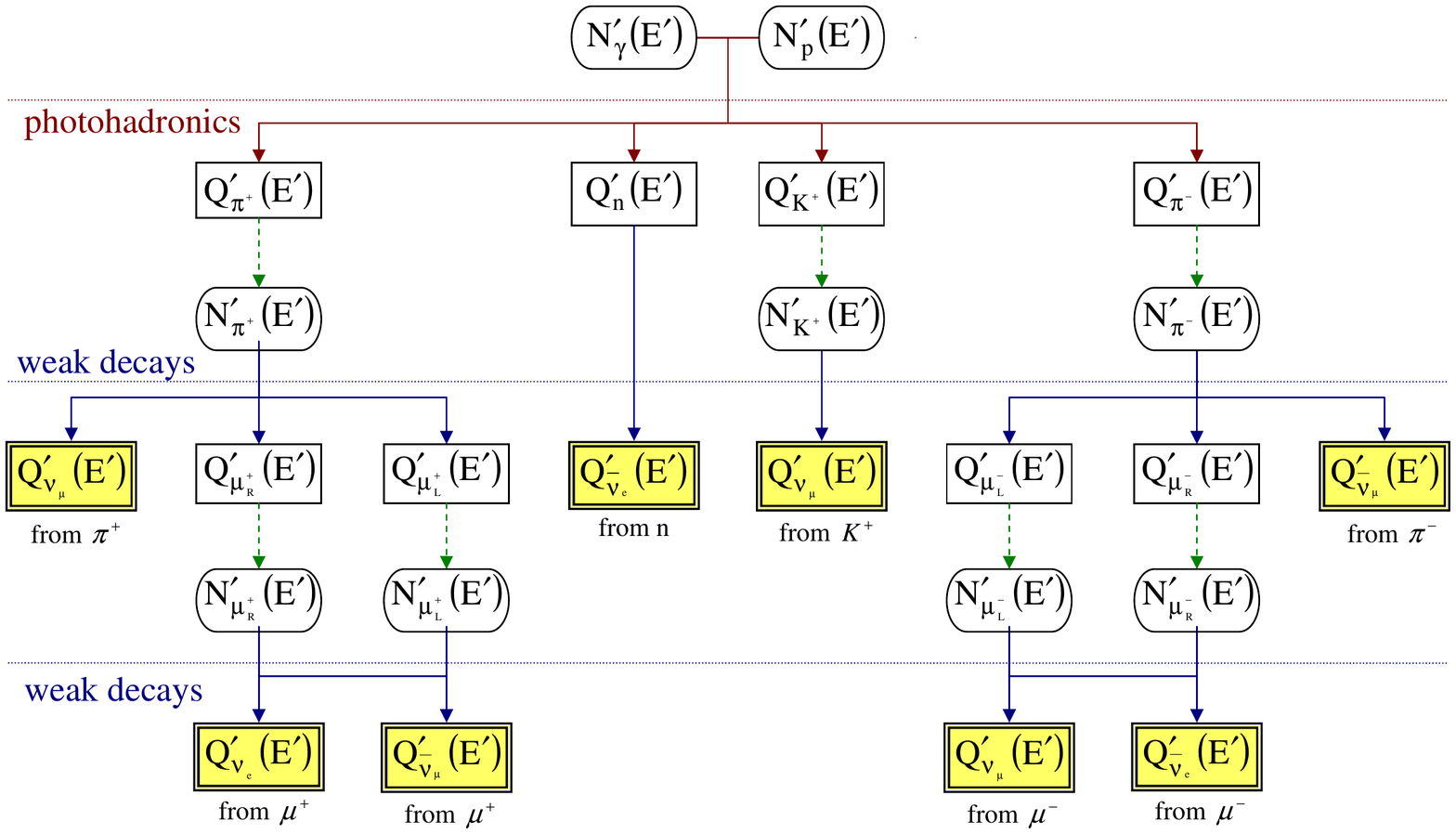}
 \mycaption{\label{fig:flowchart}Flowchart describing the model (in SRF). The functions $Q'(E)$ denote (injection) spectra per time frame $[\mathrm{\left( GeV\,cm^3\,s\right)^{-1}}]$ and $N'(E)$ steady spectra $[\mathrm{\left( GeV\,cm^3\right)^{-1}}]$ derived from the balance between injection and losses or escape. Dashed arrows stand for solving the steady state differential equation \equ{steadstate}.}
\end{figure}

To summarize, we show in  \figu{flowchart} a flowchart describing the  computation of the neutrino spectra in the SRF. Here, $Q'(E')$ (rectangular boxes) stand for (injection) spectra per time frame and $N'(E')$ (boxes with rounded corners) for steady spectra derived from the balance between injection and losses or escape. Dashed arrows stand for solving the steady state differential equation balancing energy losses and escape with injection. Weak decays and photohadronic interactions are denoted by the horizontal lines. 
The resulting neutrino spectra are marked by a double border and a colored background. Below the boxes we indicate by which parent  the neutrinos are produced. 

The transformation of the injection spectrum of the neutrinos $Q'_{\nu_\alpha}$  (in units of $\mathrm{GeV^{-1} \, cm^{-3} \, s^{-1}}$) from a single source into a point source or quasi-diffuse flux  (in units of $\mathrm{GeV^{-1} \, cm^{-2} \, s^{-1} \, [sr^{-1}]}$) at the detector $\tilde\phi_\alpha$ {\em before flavor mixing}
is given by
\begin{equation}
 \tilde\phi_\alpha = \hat N \,  \frac{(1+z)^2}{4 \pi d_L^2} \, Q'_{\nu_\alpha} \, , \qquad E_\nu=\frac{\Gamma}{1+z} \, E_\nu' \, . \label{equ:boost}
\end{equation}
Here, $\hat N$ is a (model-dependent) normalization factor depending on the volume of the interaction region and its possible Lorentz boost with respect to the observer, which will be discussed later in \Sec~\ref{sec:obs}. Here, $d_L(z)= (1+z) \, d_{\mathrm{com}}(z)$ is the luminosity distance of the source computed in standard (flat) FLRW cosmology with the values from \Ref~\cite{Komatsu:2010fb} and  $d_{\mathrm{com}}(z)$ is the comoving distance.\footnote{Note that the flux, as we define it per energy, time, and area, scales $\propto  1/d_{\mathrm{com}}^2$, where as the energy flux density, defined as energy per time frame and area, scales $\propto  1/d_L^2$.}
  From \equ{boost}, one can read off that the neutrino flux $E_\nu^2 \phi \propto 1/d_L^2$ independent of the model if only the redshift dependence is considered. In order to compute the $\phi_\beta$ ($\beta=e$, $\mu$, $\tau$) neutrino flux at the detector, we sum over all these initial neutrino fluxes of flavor $\alpha$ weighted by the usual flavor mixing
\begin{equation}
 \phi_\beta = \sum\limits_{i=1}^3 |U_{\alpha i}|^2 | U_{\beta i}|^2  \, \tilde \phi_\alpha \, .
\label{equ:flmix}
\end{equation}
We use $\sin^2 \theta_{12}=0.318$, $\sin^2 \theta_{23}=0.5$, and $\sin^2 \theta_{13}=0$ for the sake of simplicity (see, \eg, \Ref~\cite{Schwetz:2008er}; $\theta_{13}=0$ is compatible with recent T2K hint at $2.5\sigma$~\cite{Abe:2011sj}), leading to flavor equipartition between $\nu_\mu$ and $\nu_\tau$ at the detector. We sum over neutrinos and antineutrinos, \ie,  if we refer to ``$\nu_\mu$'', we mean the sum of the $\nu_\mu$ and $\bar\nu_\mu$ fluxes. 

Apart from the flux, the simplest observable may be a flavor ratio in which the detector properties are taken into account. Except from muon tracks, the event topologies currently discussed in IceCube are cascades~\cite{IcCascade:2011ui}.  If we assume that electromagnetic (from $\nu_e$) and hadronic (from $\nu_\tau$) cascades do not need to be distinguished, a useful observable is the ratio of muon tracks to
cascades~\cite{Serpico:2005sz}
\begin{eqnarray}
\rhat &\equiv& \frac{\phi_{\mu}}{\phi_{e}+\phi_{\tau}}.
 \label{equ:R}
\end{eqnarray}
 Note that neutral current events will also produce cascades and $\nu_\tau$ will also produce muon tracks in 17\% of all cases, which, in practice, have to be included as backgrounds. In \Ref~\cite{IcCascade:2011ui}, the contribution of the different flavors to the cascade rate for a $E^{-2}$ extragalactic test flux with equal contributions of all flavors at the Earth was given as: electron neutrinos 40\%, tau neutrinos 45\%, and muon neutrinos 15\% (after all cuts). This implies that charged current showers dominate and that electron and tau neutrinos are detected with comparable efficiencies, \ie,  that \equ{R} is a good first approximation to discuss flavor at a neutrino telescope.
The benefit of this flavor ratio is that the
  normalization of the source drops out. In addition, it represents the experimental flavor measurement with the simplest possible assumptions. For a pion beam source, where at the source $\nu_e:\nu_\mu:\nu_\tau = 1:2:0$, at the detector $\hat R$ is $0.5$, for a muon damped source, where at the source $\nu_e:\nu_\mu:\nu_\tau = 0:1:0$, at the detector $\hat R$ is $0.64$, and for a neutron beam source (neutrinos from neutron decays), where at the source $\nu_e:\nu_\mu:\nu_\tau = 1:0:0$, at the detector  $\hat R$ is $0.28$.

\begin{table}[t]
\begin{center}
\begin{tabular}{lllll}
\hline
Parameter & Units & Description & Typical & Ref.\\
\hline
$B'$ & G (Gauss) & Magnetic field strength & $10^3 \, \mathrm{G} \hdots 10^{6} \, \mathrm{G}$ & $300 \, \mathrm{kG}$ \\
$\Gamma$ & 1 & Lorentz boost factor & $100 \hdots 1400$ & $10^{2.5}$\\
$z$ & 1 & Redshift & $0.02 \hdots 6$ & $2$\\
\hline
$\varepsilon^{'}_{\gamma,\mathrm{break}}$ & keV & Break in photon spectrum (SRF) & $0.1 \, \mathrm{keV} \hdots 30 \, \mathrm{keV}$ & $14.8 \, \mathrm{keV}$ \\
$\alpha_\gamma$ & 1 & Photon index below break& $-1$ & $-1$ \\
$\beta_\gamma$ & 1 & Photon index above break & $-2$ & $-2$ \\
$\alpha_p$ & 1 & Proton spectral index & $-2$ & $-2$\\
\hline
$\eta$ & 1 & Acceleration efficiency & $0.1$ & $0.1$  \\
$N$ & a.u. & Normalization/luminosity & arbitrary units & \equ{wb} \\
\hline
\end{tabular}
\end{center}
\mycaption{\label{tab:params} Parameters of our model, units, description, typical values used in this study,
and a reference parameter set (chosen in \Sec~\ref{sec:typ}). }
\end{table}

Our main  model parameters are summarized in \Tab~\ref{tab:params}. In this table, the set of parameters in the last column ``Ref.'' has been chosen as one possibility to reproduce the original WB shape in the analytical version, see \App~\ref{sec:typ}. In the following, we use $E_\nu^2 \phi_\mu$ from \equ{flmix}, representative for the muon tracks in the neutrino telescope, and $\rhat$ in \equ{R}, representative for the muon track to shower ration in the neutrino telescope, as observables, \ie, we will show the results for these observables where applicable.

\subsection{Model dependence and connection to gamma-ray observations}
\label{sec:obs}

In order to demonstrate how burst to burst variations are to be interpreted in terms of different models, 
we discuss three different approaches: cannonball model-like phenomenology, a general $N$ zone fireball model with properties similar in the SRF, and conventional fireball phenomenology. In the latter two cases, we illustrate how the input parameters needed (\cf, \Tab~\ref{tab:params}) can be related to gamma-ray observations, in particular, how the photon spectrum normalization $C'_\gamma$ in \equ{targetphoton} and the proton spectrum normalization $C'_p$ in \equ{targetproton} are obtained explicitely. 

\subsubsection*{Cannonball-like phenomenology (CB)}

Suppose that the acceleration region emits isotropically in the bulk rest frame, and this whole region is ejected by the engine with a Lorentz boost $\Gamma$, such as a cannonball (CB), see  \Ref~\cite{Dar:2003vf} for a specific model.\footnote{Note that, in general, a Doppler factor depending on the viewing angle has to be used, see \eg\ \Ref~\cite{Rachen:1998fd}. }  Then the isotropic emission region will be boosted into a solid angle $\propto 1/\Gamma^2$ by simple relativistic dynamics, and the normalization factor in \equ{boost} can be written as
\begin{equation}
 \hat N = V' \, \Gamma^2 = \frac{4 \pi}{3} R'^3  \, \Gamma^2 \, .
\label{equ:boostcannon}
\end{equation}
As a consequence, $E_\nu^2 \phi \propto \Gamma^4$ if the cannonballs are otherwise alike in terms of the photon and proton spectra in the bulk rest frame, \ie, the Lorentz boost $\Gamma$ is entirely a feature of the engine.
This is, of course, a very specific assumption, which does not take into account the origin of the target photon field, as in \Ref~\cite{Dar:2003vf}.\footnote{In fact, in \Ref~\cite{Dar:2003vf} a thermal photon field which is isotropic in the source frame is assumed. Therefore, the photon energies in the SRF are proportional to $\Gamma$ (and, in addition, the Lorentz boost and Doppler factor have to be distinguished). Here, for the sake of simplicity, we assume that the target photon field is isotropic in the bulk, as for synchrotron photons, but this does not affect our argument.} Finally, note that the probability to observe such a burst will be $\propto 1/\Gamma^2$, which is determined by the probability that the observer is within the solid angle of the jet.

\subsubsection*{Fireball model with bursts alike in shock rest frame (FB-S)}

If the gamma-ray burst comes from a relativistically expanding fireball (FB), we can assume that the acceleration region expands isotropically in the source (engine) frame. This is even a good approximation if there is a jet with the opening angle $\theta \gtrsim 1/\Gamma$, since the Lorentz boost will ``overwrite'' the beaming by the Lorentz transformation into the observer's frame as long as the observer is close enough to the jet axis. Therefore, the interaction region is typically assumed to be isotropic in the source frame, and the beaming does not have an effect on the quantities we are interested in.  
On the other hand, it is implied that the probability to observe such a GRB is proportional to the solid angle of the jet $\theta^2/2$, which may be correlated with $\Gamma$.  

The normalization factor in \equ{boost} cannot simply be obtained as in \equ{boostcannon}, because the emission region is not spherically symmetric in the SRF, but in the source frame. Since it is a frequently used approach to estimate the neutrino spectrum from the gamma-ray signal, we demonstrate how this normalization is typically performed.  Because our model does not describe the neutrino production in a time-resolved way, it makes sense to relate the neutrino production to the gamma-ray fluence during the burst. The quantity of interest is the bolometric fluence $S_{\text{bol}}$ (in units of $\mathrm{erg \, cm^{-2}}$). In practice, of course, only information from specific wavelength bands may be available. It can be used to calculate the  isotropic bolometric equivalent energy $E_{\text{iso},\text{bol}}$ (in $\text{erg}$) in the source (engine) frame as 
\begin{equation}
	E_{\text{iso},\text{bol}} = \frac{4\pi \, d_L^2}{(1+\textit{z})} \; S_{\text{bol}}  \quad .
\label{equ:eisobol}
\end{equation} 
This energy corresponds to the total emitted photon energy at the source if the emission is isotropic.\footnote{In fact, for this computation, we do not need the actual (collimation corrected) emitted energy, which depends on the beaming of the jet, because any beaming effect will cancel out in our computation of the proton and photon densities.} 
It can easily be boosted into the SRF by $E'_{\text{iso},\text{bol}}= E_{\text{iso},\text{bol}}/\Gamma$.
Assuming that the emitted photons are coming from synchrotron emission of electrons (or mainly interact with electrons), the amount of energy in electrons and photons should be roughly equivalent. 
Therefore, if the photons carry a fraction $\epsilon_e$  of the total energy $E_{\text{iso},\text{tot}}$, we have
\begin{equation}
	E_{\text{iso},\text{tot}} = \epsilon_e^{-1} \cdot E_{\text{iso},\text{bol}} \, .
\label{equ:eiso}
\end{equation}
In order to compute the photon and proton densities in the SRF, it turns out to be useful to define  an ``isotropic volume'' $V_{\mathrm{iso}}$, which is the volume of the interaction region in the source frame assuming isotropic emission of the engine. Similarly, $V'_{\mathrm{iso}}$ is an equivalent volume in the SRF where only the radial direction is boosted.
Because of the intermittent nature of GRBs, the total fluence is assumed to be coming from $N$ such interaction regions.
Now one can determine the normalization of the photon spectrum in \equ{targetphoton} from
\begin{equation}
 \int  \, \varepsilon' \, N'_{\gamma}(\varepsilon') \mathrm{d}\varepsilon' = \frac{E'_{\text{iso},\text{bol}}}{N \, V'_{\text{iso}}} \quad . \label{equ:photonorm}
\end{equation}
if one assumes that the target photons actually escape from the source on the timescale considered.
Similarly, one can compute the normalization of the proton spectrum in \equ{targetproton} by
\begin{equation}
\int \, E'_p \, N'_p(E'_p) \, \mathrm{d}E'_p =  \frac{1}{f_e} \frac{E'_{\text{iso},\text{bol}}}{N \, V'_{\text{iso}}}  \, ,
\label{equ:protonorm}
\end{equation}
where $f_e$ is the ratio between energy in electrons and protons ($f_e^{-1}$: baryonic loading).
Note that in the end, we will obtain the neutrino flux $\phi$ per time frame per interaction region from \equ{boost} with $\hat N = V'_{\mathrm{iso}}$. 
Given that the magnetic field carries a fraction $\epsilon_B$ of $E_{\text{iso},\text{tot}}$, one has in addition
\begin{equation}
 U'_B = \frac{\epsilon_B}{\epsilon_e} \cdot \frac{E'_{\text{iso},\text{bol}}}{N \, V'_{\text{iso}}} \quad \text{or} \quad B' = \sqrt{8\pi \, \frac{\epsilon_B}{\epsilon_e} \cdot \frac{E'_{\text{iso},\text{bol}}}{N \, V'_{\text{iso}}}}  .
\label{equ:B}
\end{equation}
Typical values used in the literature are $f_e \sim \epsilon_e \sim \epsilon_B \simeq 0.1$ (see, \eg, \Ref~\cite{Abbasi:2009ig}).

From Eqs.~(\ref{equ:photonorm}) and (\ref{equ:protonorm}), together with \equ{boost}, one can easily read off that if different bursts have the same characteristics in the SRF, such as $N$, $V'_{\text{iso}}$, and $E'_{\text{iso},\text{bol}}$, there is no dependence of the neutrino flux $\phi$ on $\Gamma$, and $E_\nu^2 \phi \propto \Gamma^2$. As for the cannonball above, this case may be plausible if the properties in the bulk are alike, but the Lorentz boost is determined by the engine. This hypothesis is consistent with \Ref~\cite{Ghirlanda:2011bn}, assuming that the bursts are similar in the comoving frame. We will refer to it as ``FB-S'' (fireball-shock) in the following.

\subsubsection*{Conventional fireball phenomenology (FB-D)}
\label{sec:modeldep}

Here, we describe how the unknown parameters in the previous $N$ zone model are related to conventional fireball phenomenology. The gamma-ray emission is assumed to originate from the collision of internal shells of the thickness $\Delta d'$ at the collision radius $R_C$, see \Refs~\cite{Zhang:2003uk,Piran:2004ba} for reviews.
These parameters determine the size of the interaction region, \ie, in our language
\begin{equation}
	V'_{\mathrm{iso}} = 4\pi \, R_C^2 \cdot \Delta d' = 4\pi \, R_C^2 \cdot \Gamma \cdot \Delta d \quad .
	\label{equ:visoISM}
\end{equation}
The time variability of the gamma-ray signal can be described by many such collisions. In turn, the variability timescale $t_v$ of the signal at the observer can be used to draw conclusions about the geometry, leading to the well known estimates 
\begin{equation}
	R_C \simeq  2 \, \Gamma^2 \, c \, \frac{t_v}{(1+\textit{z})} \, , \quad \Delta d' \simeq \Gamma \cdot c \, \frac{t_v}{(1+\textit{z})}  = \frac{R_C}{2 \Gamma} .
	\label{equ:ism}
\end{equation}
where the effects of the cosmological redshift are taken into account.
From \equ{visoISM} we can then estimate the size of the interaction region as
\begin{equation}
	V'_{\mathrm{iso}} \simeq 4\pi \, \left(2 \, \Gamma^2 \, c \, \frac{t_v}{(1+\textit{z})}\right)^2 \cdot \left(\Gamma \cdot c \, \left( \frac{t_v}{(1+\textit{z})} \right) \right) \propto \Gamma^5 \, .
	\label{equ:visoISMexpl}
\end{equation}
In addition, if there are $N$ collisions, we can estimate that $N \simeq T_{90}/t_v$, where $T_{90}$ is the time during which 90\% of the total energy is observed. Consequently, the total neutrino fluence is obtained by multiplication with $t_v \, N = T_{90}$. \equ{visoISMexpl} implies that for fixed $t_v$, the larger $\Gamma$, the larger the interaction region, and the smaller the photon density in \equ{photonorm}, which directly enters the fraction of fireball proton energy lost into pion production $f_\pi \propto \Gamma^{-4}$~\cite{Guetta:2003wi}, or, consequently,  $E_\nu^2 \phi \propto \Gamma^{-4}$. Therefore, the main contribution to the neutrino flux is often believed to  come from bursts with small Lorentz factors, see discussions in \Refs~\cite{Guetta:2000ye,Guetta:2001cd,Guetta:2003wi}. As a consequence, recently observed bursts with high $\Gamma$'s by \textit{Fermi}-LAT would, despite their high luminosities, not lead to significant neutrino production. We will refer to this hypothesis as ``FB-D'' (fireball-detector) in the following, since the conclusions are based on the uncorrelated variation of the {\em detected} quantities.

We can now compute the magnetic field  in \equ{B} using \equ{ism} as
\begin{equation}
	B' \simeq 220 \, \left( \frac{\epsilon_B}{\epsilon_e} \right)^{\frac{1}{2}} \, \left( \frac{E_{\text{iso},\text{bol}}}{10^{53} \, \text{erg}}\right)^{\frac{1}{2}} \, \left( \frac{\Gamma}{10^{2.5}} \right)^{-3} \, \left( \frac{t_v}{0.01 \, \second} \right)^{-1} \, \left( \frac{T_{90}}{10 \, \second} \right)^{-\frac{1}{2}} \, \left( \frac{1+\textit{z}}{3} \right)^{\frac{3}{2}} \,  \kilo\text{G} 
	\label{equ:BISM}
\end{equation}
in consistency with \Refs~\cite{Guetta:2003wi,Razzaque:2006qa}.\footnote{In fact, different versions of this formula uses $L_{\gamma, \mathrm{iso}}$. However, as long as $N=T_{90}/t_v$, these results are identical apart from the redshift effect on $t_v$, which we take into account.}  This means that the reference values chosen in \Tab~\ref{tab:params}, in particular, the magnetic field, are consistent with the values of $T_{90}$, $t_v$ and $E_{\text{iso},\text{bol}}$ suggested by this formula.

\subsubsection*{Comparison among approaches}

Comparing the different approaches, CB and FB-S are attractive because variations of the bulk Lorentz factor $\Gamma$ of otherwise similar bursts in the comoving frame lead to natural descriptions of observed empirical relationships, such as the Amati and Yonetoku relationships, which are obtained from simple relativistic transformations; see, \eg, \Refs~\cite{Dar:2005gb,Ghirlanda:2011bn}. On the other hand,  FB-D is the most frequently adapted alternative in neutrino physics. From the discussion above, it is clear that the conclusion that small Lorentz factors dominate the aggregated neutrino flux only holds in FB-D, based on the geometry estimate in \equ{visoISMexpl}. However, the exact shell widths are unknown. More importantly, this conclusion implies that $t_v$ is held fixed and therefore considered as input. However, $t_v$ is a not even well-defined observable. For example, from the discussion in \Ref~\cite{Gao:2011ej}, it is clear that there can be several superimposed variability components. In addition, the variability depends on the wave length band and the fastest variation may not even be observed because of the instrumental cutoff.\footnote{See, \eg, Figs.~2 and~3 in  \Ref~\cite{Gao:2011ej}, where in some cases the a dip seems to be present below the cut-off frequency.} 
On the other hand, the alternative FB-S corresponds to the very recently found correlations in \Ref~\cite{Ghirlanda:2011bn},  which translates into $E_\nu^2 \phi_\nu \propto \Gamma^2$. 
Of course, there may be additional interesting correlations, such as $\Gamma \propto t_\nu^{-3/5}$ for fixed $V'_{\mathrm{iso}}$; \cf, \equ{visoISMexpl}.  Since it is not clear which parameters are to be considered as input (the properties at the source or at the observer, such as $t_v$), we will test three hypotheses: $E_\nu^2 \phi_\nu \propto \Gamma^4$ (CB),  $E_\nu^2 \phi_\nu \propto \Gamma^2$ (FB-S; parameters in SRF fixed),  and $E_\nu^2 \phi_\nu \propto \Gamma^{-4}$ (FB-D; parameters at detector fixed). The truth may lie somewhere in between, but will require the further test of empirical correlations.

\section{Diffuse neutrino fluxes}
\label{sec:diffuse}

Diffuse neutrino fluxes are generated by many bursts, which are assumed to be not resolvable individually. On the other hand, stacking analyses are based on a ``finite'' number of bursts, typically a few hundred bursts observed in gamma-rays. From a stacking limit, a quasi-diffuse limit can be computed if the total number of bursts per year is roughly known. This limit will obtain a systematical error coming from the statistics of the stacking sample, see  \Sec~\ref{sec:lowstatistics}. Note that there may be also selection effects\footnote{One example for such a selection effect would be the threshold effect for the minimal detectable flux, which is important for the detection in photons. In \App~\ref{sec:threshold} we briefly discuss the consequence of this selection effect.} or uncertainties in the total number of bursts, which we do not discuss in this work. In practice, it is assumed that $\mathcal{O}(1000)$ GRBs per year happen in the observable universe, see, \eg~, \Ref~\cite{Becker:2005ej}. Therefore, even a diffuse flux is based on a finite number of GRBs, which are, for ten years of operation of the neutrino telescope, $\mathcal{O}(10 \, 000)$ GRBs contributing to the diffuse flux. Therefore, we use the same method to compute diffuse and quasi-diffuse fluxes:
we generate a probability distribution for one parameter, say redshift $z$, and compute $n$ neutrino fluxes
with parameters drawn randomly from this distribution. The aggregated neutrino flux is then summed
over these $n$ individual burst fluxes, and, after that, normalized to our reference flux described by Eqs.~(\ref{equ:wb}) and (\ref{equ:WB}) in terms of the total energy flux of the  neutrinos. Note that this method takes into account the relative contribution from the individual fluxes, but the absolute normalization of the aggregated flux is not an outcome of our approach; it will be discussed elsewhere~\cite{Hummer:prep}. 
For the sake of simplicity, we do not distinguish between short and long bursts in this analysis, while we choose typical parameters of long bursts.

The method has two advantages: First of all, diffuse and quasi-diffuse fluxes are computed with the same method, which means that, by varying $n$, the borderline between diffuse and quasi-diffuse limits can be illustrated, and low statistics effects can be quantified, see \Sec~\ref{sec:lowstatistics}. Second, it may be more realistic than a pure integration of the flux, especially if there are observable effects between $\mathcal{O}(10 \, 000)$ and infinitely many bursts contributing to the diffuse flux. As we will demonstrate later, $n = 10 \, 000$, which we use throughout this section, is already a very good approximation for the diffuse limit. But this is an outcome of our analysis, not an input assumption.  The results in this section are also representative for the integrated flux from a stacking analysis with very high statistics. Note, however, that the neutrino effective area of the neutrino telescope may lead to selection effects of individual bursts, which we do not discuss in this work.

Since each of the parameters, as  we will discuss in the following, has different theoretical issues to be addressed, we vary only one parameter at the time for clarity, and keep the others fixed. We consider the main input parameters for our model, see \Tab~\ref{tab:params}: $z$, $B'$, $\Gamma$, and the photon spectral shape. 
For each parameter, there are mainly two challenges:
\begin{enumerate}
 \item
  The description of the (theoretical) distribution of the parameter.
 \item
  The interpretation of a different value of the parameter in terms of the relative contribution to the total flux.
\end{enumerate}
For example, the redshift $z$ is a relatively simple case: The distribution is typically assumed to follow the star formation rate in some form, and the consequences of a different parameter value are independent of the source model if the GRBs are otherwise equivalent at the source. A very different example is $\Gamma$ (see discussion in \Sec~\ref{sec:modeldep}): The theoretical distribution of the parameter is not directly known, and the consequences of a different value of $\Gamma$ depend on the assumptions for the model. Therefore, it is clear that variations of $z$ and $\Gamma$ cannot be compared at the qualitative level. Before we vary a parameter, we also show the consequences of different values explicitely.

\subsection{Redshift}
\label{sec:redshift}

\begin{figure}[t!]
\begin{center}
\includegraphics[width=\textwidth]{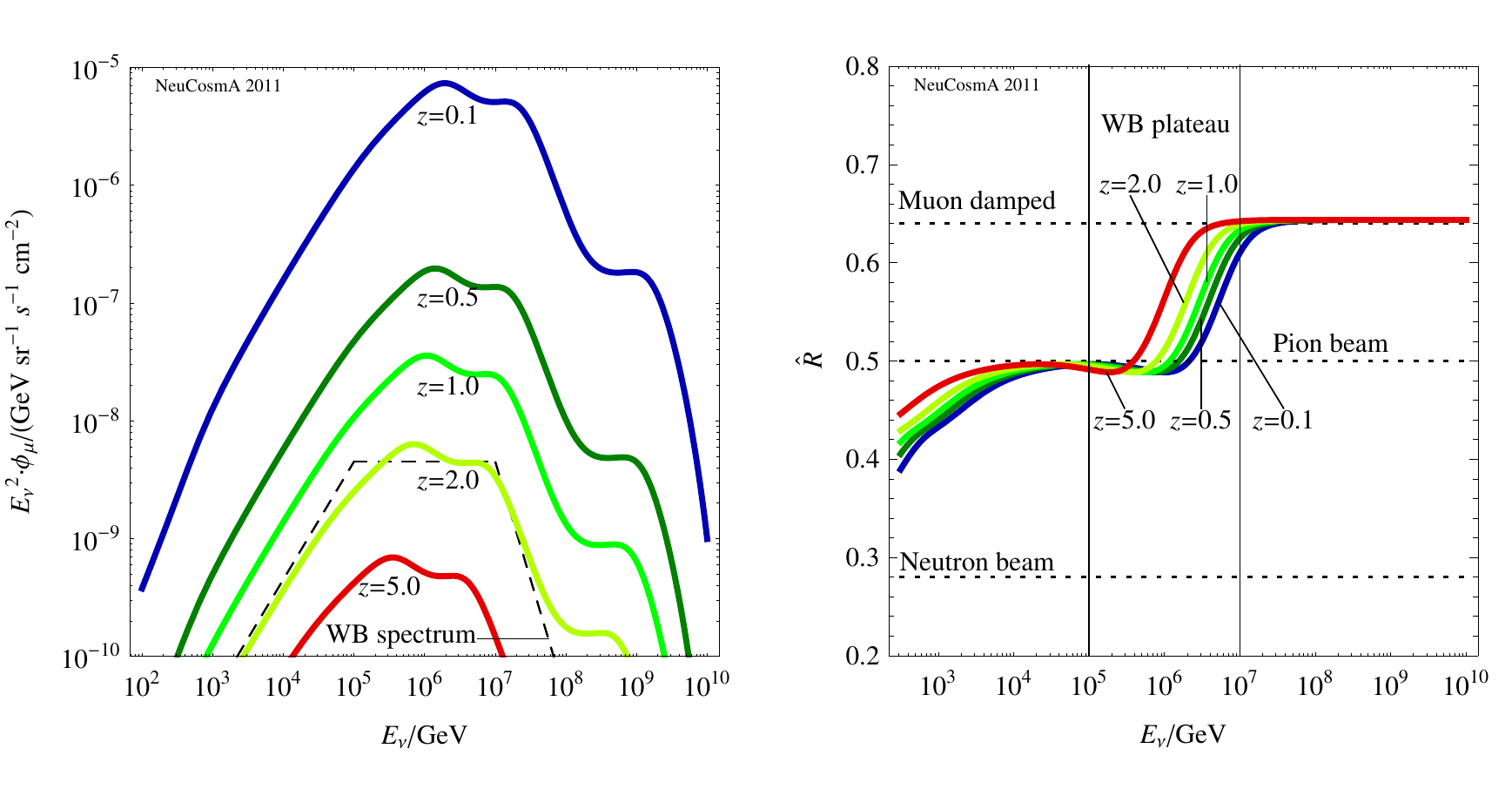}
\end{center}
\mycaption{\label{fig:redshiftdep} Neutrino flux (left) and flavor ratio (right) for the reference values from \Tab~\ref{tab:params}, where the redshift \textit{z} is varied. The burst with $z=2$ is normalized to the WB spectrum, the other bursts scale with redshift as given by \equ{boost}. }
\end{figure}

Here, we assume that all GRBs are alike at the source and the different fluxes and spectra are generated solely by the different redshifts of the bursts. We show in \figu{redshiftdep} the impact of the redshift on the neutrino flux (left) and flavor ratio (right) for the GRB with the reference values from \Tab~\ref{tab:params}, with \textit{z} varied.  From the discussion after \equ{boost}, we have $E_\nu^2 \phi_\nu \propto d_L^{-2}$, \ie, bursts with small redshifts contribute most. Since the energy $E_\nu \propto (1+\textit{z})^{-1}$, there is hardly any shift of the spectrum, as it can best be seen in the right panel. Nevertheless, it is not clear if the contributions from different bursts will lead to some averaging as long as the distribution of bursts is not considered.

\begin{figure}[t!]
\begin{center}
\includegraphics[width=\textwidth]{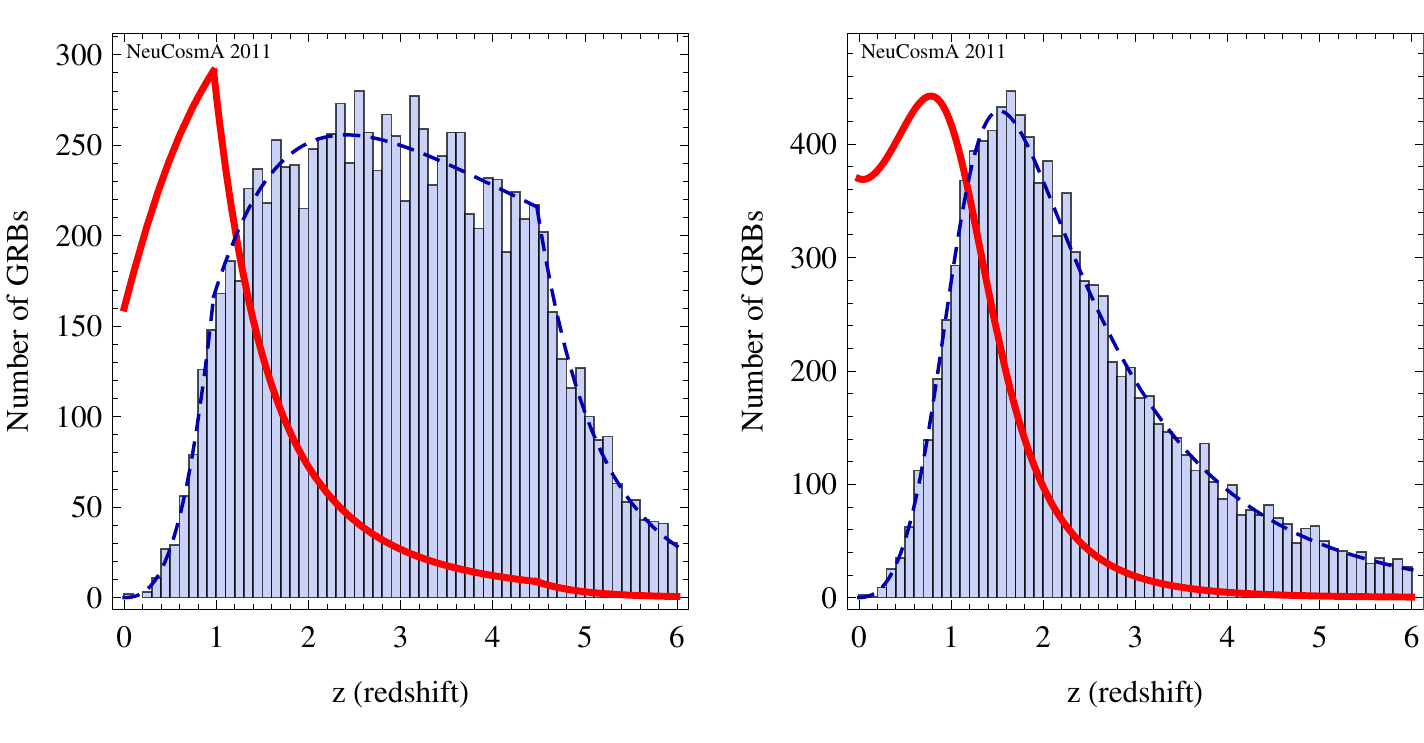} 
\end{center}
\mycaption{\label{fig:redshiftdistr} Distribution of $10 \, 000$ bursts $\mathrm{d}\dot{N}/\mathrm{d}\textit{z}$ as a function of redshift  (histograms) and relative contribution of the individual GRBs $d_L^{-2} \, \mathrm{d}\dot{N}/\mathrm{d}\textit{z}$ (solid curves). The dashed curves show the exact distribution functions.
The left panel uses the star formation rate from Hopkins and Beacom~\cite{Hopkins:2006bw} with the  correction $\mathcal{E}(z)$ from Kistler et al.~\cite{Kistler:2009mv}, the right panel shows the ``star formation rate~1'' from Porciani \& Madau~\cite{Porciani:2000ag} without correction. }
\end{figure}

Since long GRBs are assumed to originate from the death of massive stars, it is natural to assume that the GRB rate traces the star formation rate. For example, Hopkins and Beacom use a piecewise steady approach to parameterize the star formation rate density (per comoving volume) as~\cite{Hopkins:2006bw}:
\begin{equation}
	\dot{\rho}_*(z) \propto \left\{ \begin{array}{ll} (1+\textit{z})^{3.44} & 0 < \textit{z} \leq 0.97 \\ 10^{1.09} \cdot (1+\textit{z})^{-0.26} & 0.97 < \textit{z} \leq 4.48 \\ 10^{6.66} \cdot (1+\textit{z})^{-7.8} & 4.48 < \textit{z}  \le 6 \end{array} \right. \quad .
	\label{equ:HBSFR}
\end{equation}
As a maximum value for the redshift, Hopkins and Beacom use a cut-off at $\textit{z}_{\text{max}} = 6$ because of the low statistics at higher $z$. Furthermore, we use $\textit{z}_{\text{min}} \simeq 0.02$ from the scale the universe becomes homogeneous (about 100~Mpc). We assume that the redshift distribution of the bursts follows (see, \eg, \Ref~\cite{Kistler:2009mv})
\begin{equation}
	\frac{\mathrm{d}\dot{N}}{\mathrm{d}\textit{z}} \propto \mathcal{E}(\textit{z}) \, \dot{\rho}_*(\textit{z})\, \frac{\mathrm{d}V/\mathrm{d}\textit{z}}{1+\textit{z}} \quad .
	\label{equ:Burstrateredshift}
\end{equation}
Here, we have used the comoving volume per unit redshift $\mathrm{d}V/\mathrm{d}\textit{z} \, (1+\textit{z})^{-1}$, where the factor $1+z$ accounts for the cosmic time dilation. In addition,
we use the correction  $\mathcal{E}(\textit{z}) = \mathcal{E}_0 (1+\textit{z})^{1.2}$ introduced in \Ref~\cite{Kistler:2009mv} to account for the fraction of stars resulting in GRBs (strong evolution case). We show the distribution of 10000 bursts in redshift in \figu{redshiftdistr}, left panel, for  the distribution in \equ{Burstrateredshift} as histogram. For comparison to a very different model, the distribution for the model ``SF1''  by Porciani and Madau~\cite{Porciani:2000ag} (without correction factor, \ie, star formation rate evolution) is shown in the right panel of \figu{redshiftdistr}. 

The ``typical value'' for the redshift of a GRB according to the star formation rate in \equ{HBSFR} is roughly around $\textit{z} = 1 - 2$. With the correction factor $\mathcal{E}(\textit{z})$ these values are shifted to higher redshifts with a broad peak around $\textit{z} = 1 - 5$ (left histogram). In the right histogram (without correction factor), the peak is at about  $\textit{z} = 1 - 3$.
However, do the GRBs around these values also contribute most to the neutrino flux? For this question, we show in both panels of \figu{redshiftdistr} the function $1/d_L^2 \, \mathrm{d}\dot{N}/\mathrm{d}\textit{z}$ measuring the relative contribution of individual bursts, which includes both the redshift distribution function $\mathrm{d}\dot{N}/\mathrm{d}\textit{z}$ and the weight function  $1/d_L(z)^2$ scaling $E_\nu^2 \phi$ (\cf, discussion after \equ{boost}). Clearly, the weight function prefers smaller values of $z$, whereas the  redshift distribution pulls to larger values, leading to a local narrow maximum at $z=1$. From this generic argument, we conclude that the ``typical GRB'' dominating the neutrino flux has $z \sim 1$, rather than larger values of $z$ often used in the literature. If the redshift of a GRB is not measured, it may therefore be the best estimate for the neutrino flux computation. For instance, choosing $z=2$ or $z=3$ for the unknown redshift may, depending on the method, lead to an overestimate of the bolometric equivalent energy and therefore neutrino production; \cf, \equ{eisobol}.\footnote{In fact, the IceCube stacking method in \Refs~\cite{Abbasi:2009ig,Abbasi:2011qc} does not compute the isotropic equivalent luminosity, needed for the pion production efficiency, back from the observation on a burst-by-burst basis, but uses instead a fixed value for the isotropic equivalent luminosity for the pion production. Therefore, mainly the break energies of the neutrino spectrum are affected by a different standard value of $z$.}
We have checked that this conclusion does not depend on the star formation rate or the correction factor $\mathcal{E}(\textit{z})$, even if one of the other star formation rates given in \Ref~\cite{Porciani:2000ag} is used.\footnote{We tested this claim by calculating a correctional factor $\mathcal{E}(\textit{z})$ for the SFR from \Ref~\cite{Porciani:2000ag} analogous to the approach from \Ref~\cite{Kistler:2009mv}. Even though this factor leads to more bursts at higher redshifts, the peak contribution still comes from bursts at $z \simeq 1$. } Therefore, we only use the distribution in the left panel of \figu{redshiftdistr} in the following. It is also noteworthy that at around $z=1$ the number of contributing bursts is quite low (compare to histogram), which means that strong fluctuations may be expected in low statistics samples.

\begin{figure}[t!]
\begin{center}
\includegraphics[width=\textwidth]{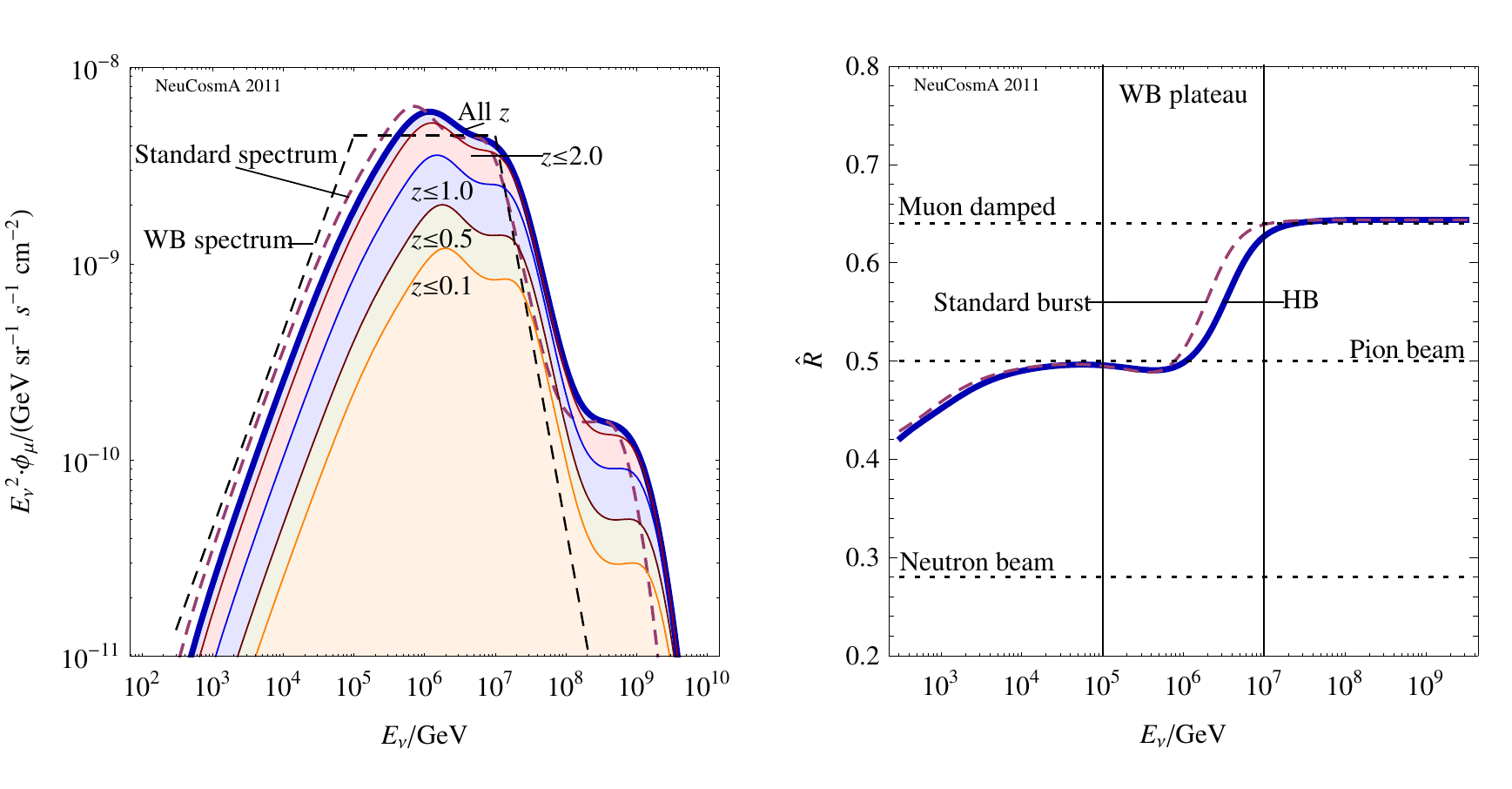}
\end{center}
\mycaption{\label{fig:redshiftresult} Neutrino flux (left) and flavor ratio (right) for the redshift distribution according to Hopkins and Beacom (HB), as depicted in \figu{redshiftdistr}, left panel. The thick solid curves show the integrated flux, the thin solid curves the contributions from individual redshift ranges (left panel only). The dashed curves show the reference values from \Tab~\ref{tab:params}. The thin dashed curve in the left panel is the WB reference, the total fluxes are normalized to.}
\end{figure}

We show in \figu{redshiftresult} the neutrino flux (left) and flavor ratio (right) for the quasi-diffuse flux. The thick solid curves show the integrated flux, the thin solid curves the contributions from individual redshift ranges (left panel only). The dashed curves show the reference values from \Tab~\ref{tab:params}. Obviously, the diffuse flux peaks at somewhat higher energies than the WB reference, because of the main contribution to the flux coming from smaller redshifts.  The same effect is visible for the flavor ratio. However, neither the characteristic wiggles in the shape, nor the flavor ratio transition from pion beam to muon damped source are averaged out. The reason is the relatively sharp peak in the weight functions in \figu{redshiftdistr}, which hardly depends on the star formation rate. One can see from the relative contribution curves in \figu{redshiftresult} (left panel) that indeed $z \simeq 1$ dominates the flux. 
The weak dependence on the star formation rate is consistent with \Ref~\cite{Murase:2005hy}. This conclusion obviously changes if there are contributions with a different systematics, such as from GRBs originating from population~III stars~\cite{Gao:2011jt}, which have different population characteristics.

\subsection{Magnetic field}

\begin{figure}[t!]
\begin{center}
\includegraphics[width=\textwidth]{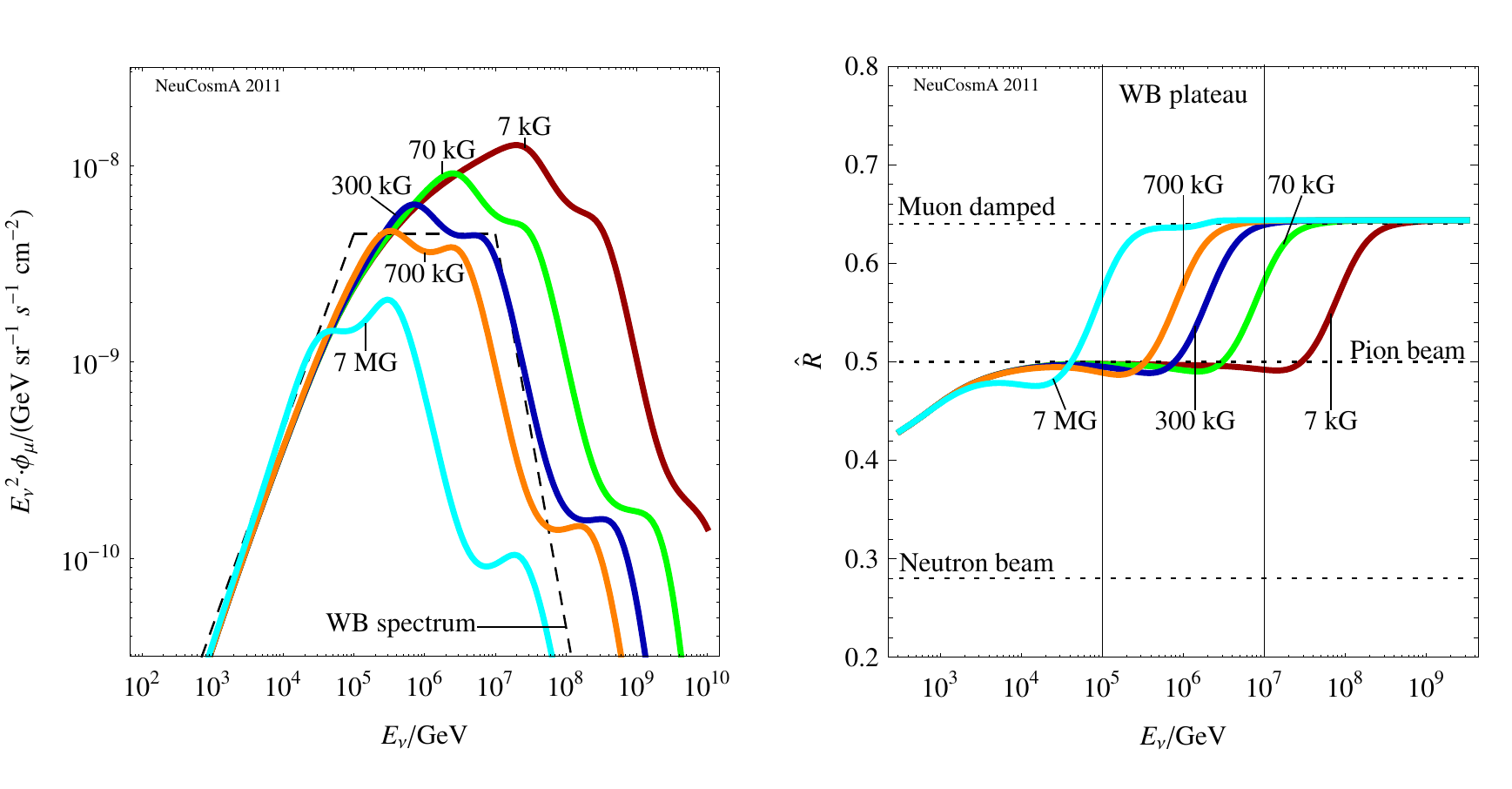}
\end{center}
\mycaption{\label{fig:Bdep} Neutrino flux (left) and flavor ratio (right) for the reference values from \Tab~\ref{tab:params}, where the magnetic field $B'$ is varied. The burst with $B'=300 \, \text{kG}$ is normalized to the WB spectrum, leading to normalization factors $C'_p$ and $C'_{\gamma}$ for the input proton and photon spectra. The same proton and photon spectra, including normalization factors, are used for the other bursts. }
\end{figure}

Assuming that the magnetic field $B'$ does not significantly change the energy budget of the jet, it will only have a significant impact on the maximal proton energy and the cooling of the secondaries. We show the effect of different values of $B'$ in \figu{Bdep}.  As can be seen from the different curves, the flux shape changes due to the magnetic field dependence of the synchrotron cooling breaks, see \equ{ec}. More precisely, higher magnetic fields lead to synchrotron losses dominating at lower energies. Therefore, fewer neutrinos are produced in case of high magnetic fields  compared to lower magnetic fields. In addition, for $B^{'} = 7 \, \text{MG}$, the highest peak is no longer determined from the muon break, but the pion break. As can be seen from the right plot in \figu{Bdep}, the effect on the flavor ratio $\rhat$ is qualitatively different from the effect on the flux shape: depending on the value of $B'$, the transition of the flavor ratio from pion beam source to muon damped source shifts in energy, but the flavor composition does not change. This shift is also a consequence of the breaks due to synchrotron cooling; the higher the magnetic field, the lower the energy of the transition. Therefore, it is obvious that the flavor ratios may be a direct handle on the magnetic field inside the source. Note that for $B^{'} = 7 \, \text{MG}$, the small dip in the flavor ratio at low energies is the result of a pile-up effect.

\begin{figure}[t!]
\begin{center}
\includegraphics[width=\textwidth]{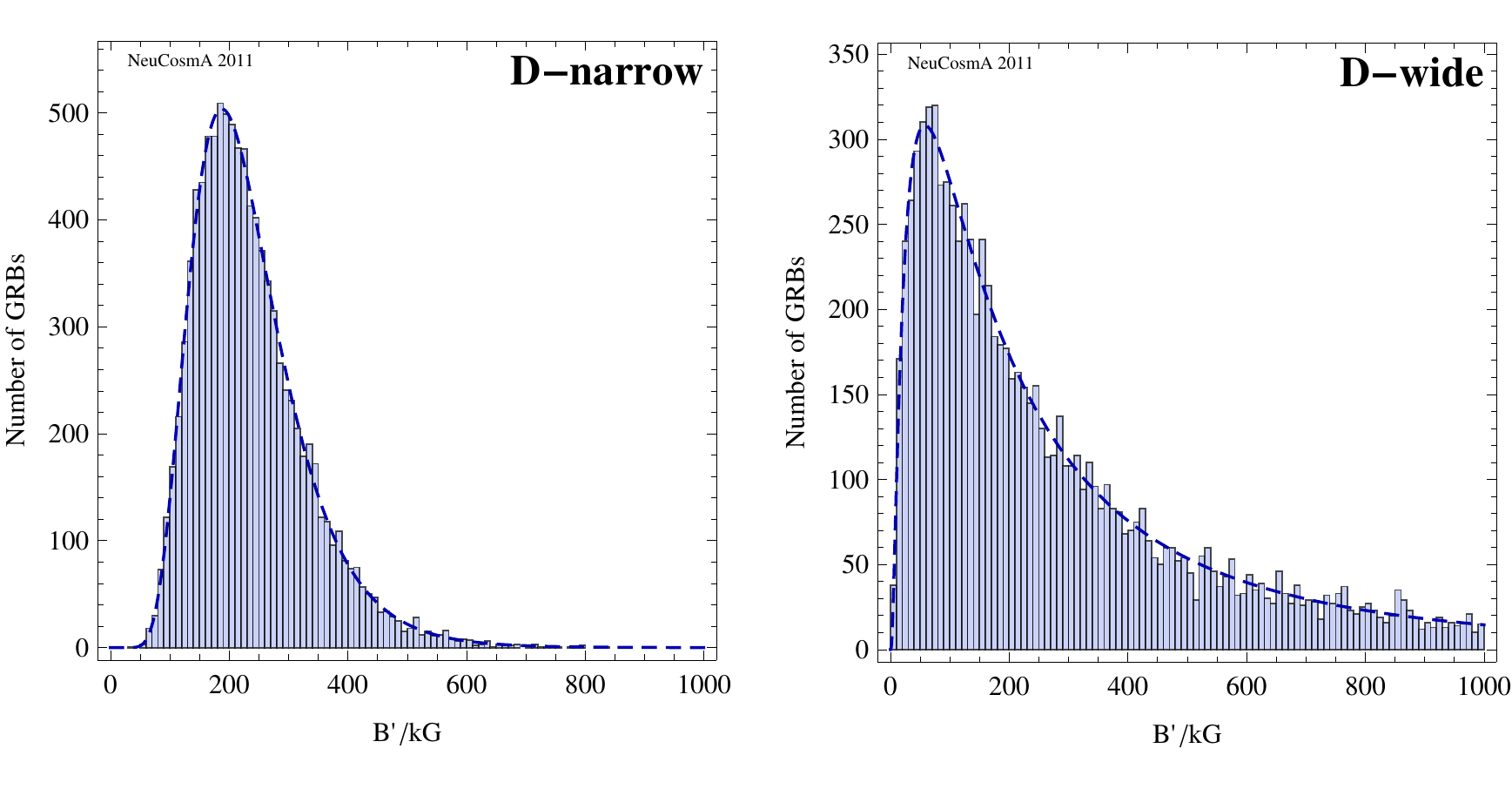}
\end{center}
\mycaption{\label{fig:Bdistr} Distribution of $10 \, 000$ bursts as a function of the magnetic field $B'$ (histograms). The left panel uses the log-linear distribution with the $3\sigma$ range of the exponent going from $70 \, \mathrm{kG} \lesssim B' \lesssim 700 \, \mathrm{kG}$, the right panel shows the broader version with the range $B' = 7 \, \mathrm{kG} \hdots 7 \, \mathrm{MG}$, see main text for details.}
\end{figure}

For the magnetic field, the main problem is the theoretical distribution of the values of $B'$, which is unknown. First of all, note that the value of $B'$ enters linearly in \equ{ec}, describing the spectral break from the energy losses. On the other hand, it depends only on the square root of the energy budget, see \equ{B}. Therefore, strong variations of $B'$ will drastically affect the spectral shape, whereas the energy budget is hardly affected. In order to get an educated guess for the distribution of $B'$, one may vary $\epsilon_B/\epsilon_e$ from $0.1$ to $10$ with a normal logarithmic distribution, leading to $70 \, \mathrm{kG} \lesssim B' \lesssim 700 \, \mathrm{kG}$ in the FB-S (or CB) case.  In the FB-D case, one can easily see from \equ{BISM} that variations of $E_{\mathrm{iso,bol}}$, $t_v$, $T_{90}$, and $z$  also have only a moderate impact on $B'$, whereas $\Gamma$ enters to third power. For example, for $\Gamma = 100 \hdots 1000$, we have $B' = 7 \, \mathrm{kG} \hdots 7 \, \mathrm{MG}$, suggesting a broader distribution of $B'$. 
This, however, depends on the view point: it is correct if the observables in \equ{BISM} are used as input, and the quantities are calculated back to the SRF. However, if the bursts instead share similar characteristics in the SRF, such as $E'_{\mathrm{iso, bol}}$, $N$, and $V'_{\mathrm{iso}}$, see \equ{B}, or even $E_{\mathrm{iso, bol}}$, $R_C$, and $\Delta d'$ (these parameter sets are related by $\Gamma$, but not $B'$),  the narrower distribution is more suggestive. We therefore choose two distributions, which are shown in \figu{Bdistr} and defined as  
\begin{eqnarray}
	 B' &=& 10^x \, \text{G} \notag \\
	\text{with} \quad P(x) &\propto& e^{-\frac{\left( x-\bar{x}  \right)^2}{2 \, \sigma^2}} \quad .
\end{eqnarray}
Here, $\bar{x}$ is the center of the peak in $x$, and $\sigma$ is the standard deviation of the Gaussian. For our simulations we used a central value $\bar{x} = 5.34$, which corresponds to $B' \simeq 289 \, \text{kG} \; (\sim 300 \, \text{kG})$. The standard deviation for the first distribution (narrower distribution, left plot of \figu{Bdistr}) is $\sigma_1 = 0.17$, while the standard deviation for the second (wider distribution, right plot of \figu{Bdistr}) is $\sigma_2 = 0.50$.\footnote{Note that the Gaussian distribution of the exponent leads to a slightly different distribution on a linear scale. Most notably, the peak is no longer at the median value of the Gaussian distribution, but at a smaller value. This comes from the conservation of probabilities during the change from logarithmic to linear scale and leads to an additional suppression of high $B'$ values with $B'^{-1}$ on a linear scale, see \figu{Bdistr}. In addition, note that we assume that a change of the magnetic field does not effect the amount of energy present in protons and photons. However, if the total energy was fixed, a higher magnetic field would also shift the energy fractions, leading to less energy available for protons and photons, and therefore even less neutrino production.}

\begin{figure}[t!]
\begin{center}
\includegraphics[width=\textwidth]{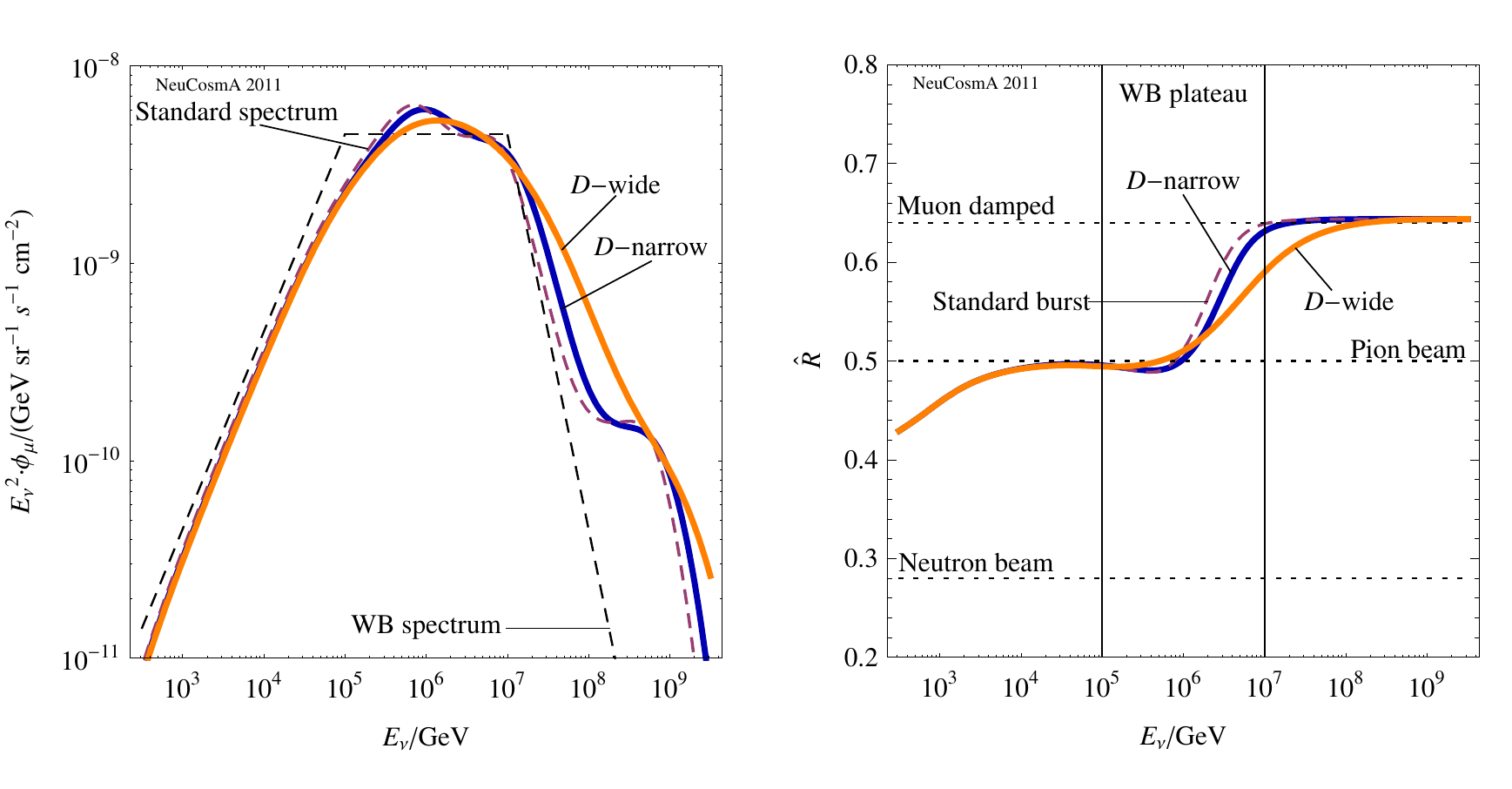}
\end{center}
\mycaption{\label{fig:Bresult} Neutrino flux (left) and flavor ratio (right) for the magnetic field distributions in \figu{Bdistr}. The thick solid curves show the flux for the two magnetic field distributions. For comparison, we show the reference burst with values from \Tab~\ref{tab:params} as a dashed curve. The thin dashed curve in the left panel is the WB reference, the total fluxes are normalized to.}
\end{figure}

We show the diffuse flux from $10 \, 000$ individual bursts in \figu{Bresult}, using the magnetic field distributions in \figu{Bdistr}. In both panels, we can see that the results for the narrow distribution (D-narrow) do not deviate much from the results of a single burst (dashed curves). The double peak and the contribution from kaons are clearly visible and deviate only slightly from the standard burst. However, in case of the broad distribution (D-wide), the flux shape (left panel) is washed out to a single peak with the maximum at around $10^6 \, \giga\electronvolt$. Neither  double peak and kaon contribution are visible anymore, which comes from the different break energies for the different bursts because they depend on the magnetic field $B'$. Hence, the neutrino flux below the (lowest) muon break is unaffected by the magnetic field distribution.
For the flavor ratio (right panel), the transition from pion beam to muon damped source occurs over a  wider energy range for the distribution ``D-wide'' than for ``D-narrow''. As a consequence, $\rhat$ carries two pieces of information:  the (mean) magnetic field and the broadness of the magnetic field distribution, while being hardly affected by any other astrophysical parameter, radiative process, \etc\ (apart from $\Gamma$, which shifts the energy of the transition).  This information can then be translated into clues for the models, such as  how alike the bursts are actually in the SRF.

\subsection{Lorentz boost}
\label{sec:lorentz}

As we have discussed in \Sec~\ref{sec:modeldep}, one of the main issues for the Lorentz boost $\Gamma$ is the translation of $\Gamma$ into the relative contributions of the bursts, which is model-dependent. Here, we test three different hypotheses:
\begin{description}
\item[CB]
  Cannonball-like model, bursts alike in SRF.  As a consequence, $E_\nu^2 \phi_\nu \propto \Gamma^4$.
  The probability to observe a burst is $\propto 1/\Gamma^2$.
\item[FB-S (Fireball-Shock)]
  Fireball model, bursts alike in SRF. As a consequence, $E_\nu^2 \phi_\nu \propto \Gamma^2$.
  The probability to observe a burst is $\propto \theta^2$ ($\theta$: jet opening angle).
\item[FB-D (Fireball-Detector)]
  Fireball model, bursts alike at the detector. As a consequence, $E_\nu^2 \phi_\nu \propto \Gamma^{-4}$.
 The probability to observe a burst is $\propto \theta^2$.
\end{description}
These three interpretations may include the most extreme cases, which means that the actual consequences of different values of $\Gamma$ may lie somewhere in the middle.  Note that assumptions close to  hypothesis FB-D have been typically implied in the literature, such as \Refs~\cite{Halzen:1999xc,AlvarezMuniz:2000st,Gupta:2006jm}, whereas \Ref~\cite{Ghirlanda:2011bn} points towards FB-S

\begin{figure}[tp]
\begin{center}
\includegraphics[width=\textwidth]{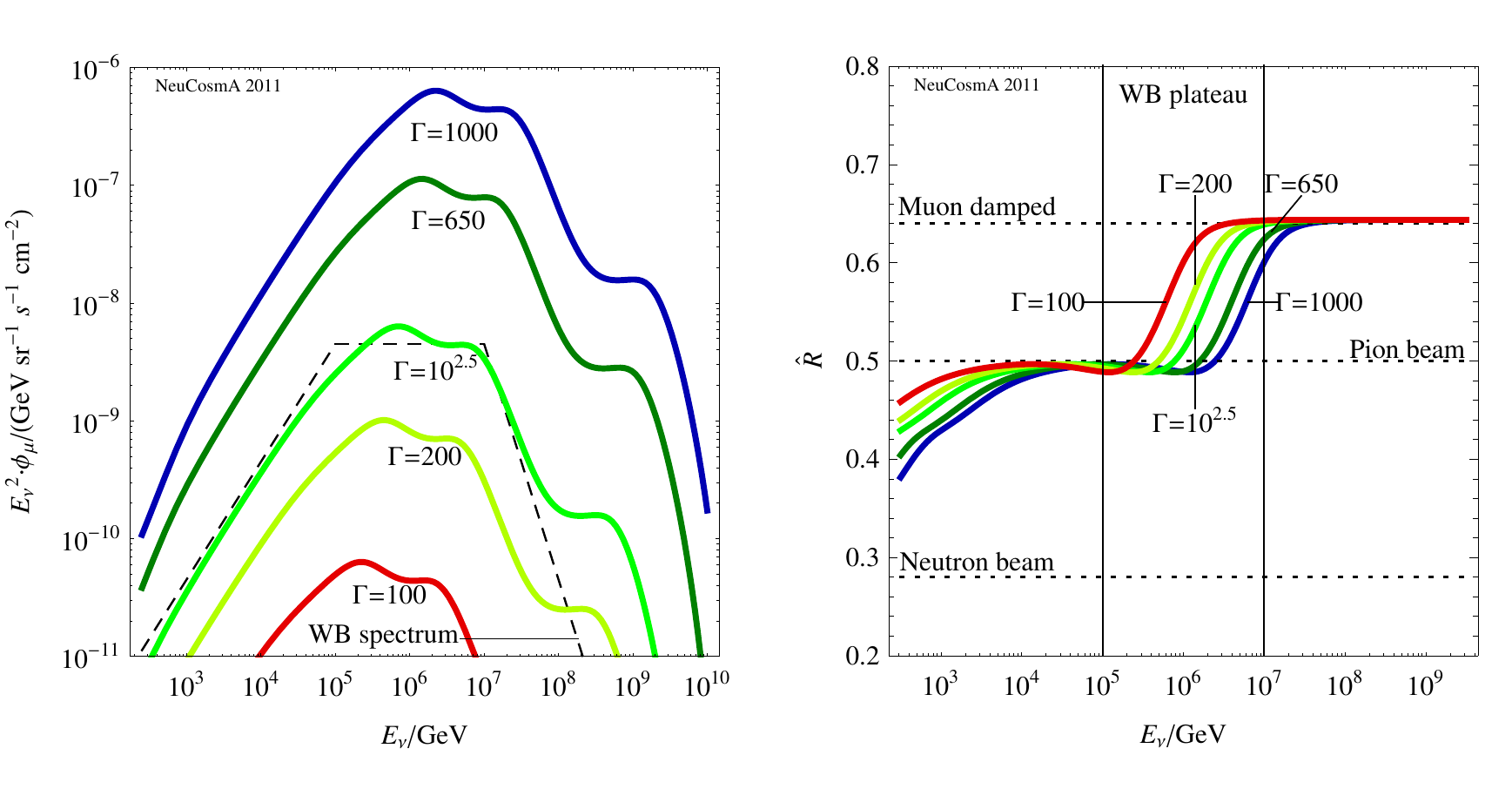}
\end{center}
\mycaption{\label{fig:gammadep} Neutrino flux (left) and flavor ratio (right) for the reference values from \Tab~\ref{tab:params}, where the Lorentz boost $\Gamma$ is varied. The burst with $\Gamma=10^{2.5}$ represents our standard burst as in \Tab~\ref{tab:params} and is normalized to the WB spectrum. In the shown figures the  hypothesis ``CB'' is used, which determines the scaling of the flux with $\Gamma$, \ie, the proton and photon spectra are normalized to the standard burst, and only $\Gamma$ is varied among the bursts.}
\end{figure}

We show in \figu{gammadep} the impact of $\Gamma$ on the neutrino flux (left) and flavor ratio (right) for the GRB with the reference values from \Tab~\ref{tab:params} for hypothesis~CB. In the shown plots we have five different values to illustrate the effect of the Lorentz factor. The curve with $\Gamma = 10^{2.5}$ represents our standard burst from \Tab~\ref{tab:params}, while the values of $\Gamma = 100$, and $1000$ are commonly used minimal and maximal values -- although \textit{Fermi}-LAT has observed GRBs with higher values of $\Gamma$, such as for GRB 080916C. From  \equ{boost}, it is clear that bursts with large Lorentz factors dominate, which also shift the flavor ratio transition from pion beam to muon damped source to higher energies. The strong Lorentz factor dependence comes from the Lorentz boost of the energy and the forward collimation of the Lorentz cone for CB: the opening angle is proportional to $1/\Gamma^2$. 

\begin{figure}[tp]
\begin{center}
\includegraphics[width=\textwidth]{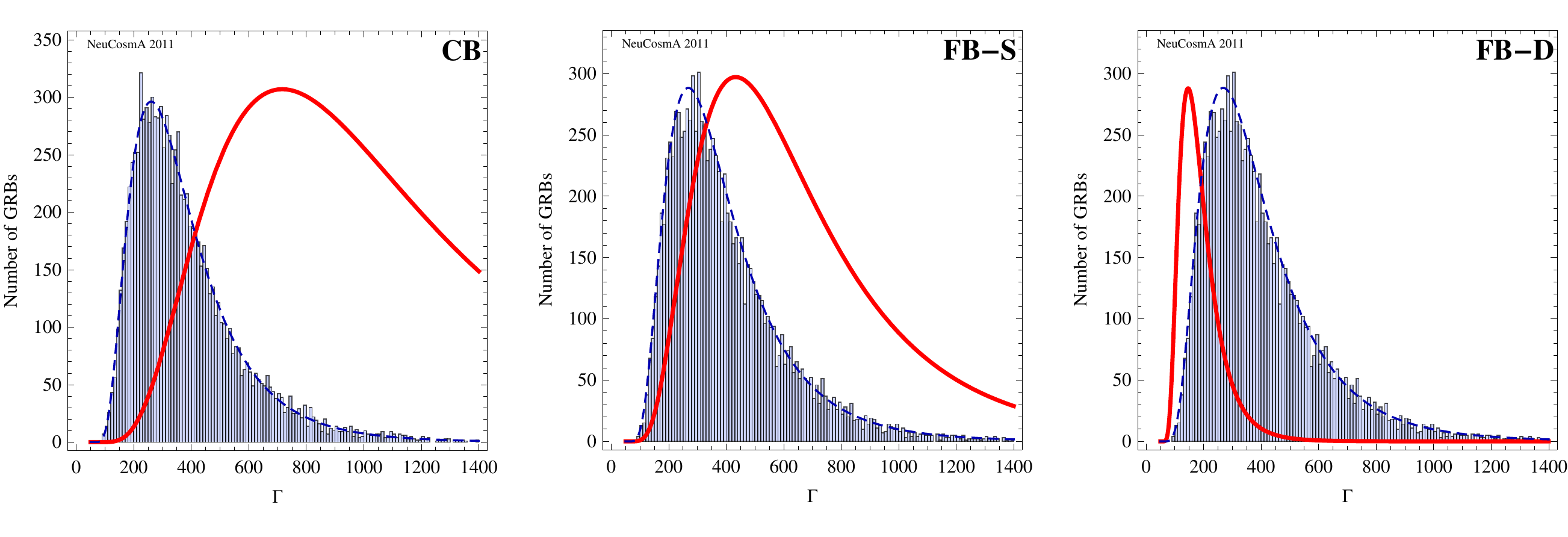}
\end{center}
\mycaption{\label{fig:Gammadistr} Distribution of $10 \, 000$ bursts as a function of the Lorentz factor $\Gamma$ (histograms) and the relative contribution of a value of $\Gamma$ to the total diffuse flux (thick solid curves). The relative contribution consists of the distribution of bursts and a scaling factor depending on the hypothesis (see main text). Additionally, the dashed thin curves represent the analytic version of the distribution of the number of bursts. The different panels correspond to the different hypotheses, as explained in the main text.}
\end{figure}

Although the theoretical distribution of $\Gamma$ is unknown, there exist empirical analyses. For the sake of simplicity, we assume that it should roughly reproduce the observed (derived) distribution in \Ref~\cite{Guetta:2003wi} (\cf, Fig.~1b).\footnote{Note that this analysis is based on the maximally observed photon energy, whereas different approaches, such as based on the afterglow, may be more reliable; see, \eg, \Ref~\cite{Liang:2009zi}. However, while the peak contribution in $\Gamma$ may be somewhat affected by the distribution function, the qualitative results from this section depend rather on the tested hypothesis, \ie, interpretation of different values of $\Gamma$.} Therefore, we empirically use a probability distribution 
\begin{equation}
	P(\Gamma) \propto \frac{1}{\Gamma^n} \, e^{-\frac{\left( \log_{10}(\Gamma-50) - \bar{x} \right)^2}{2 \, \sigma^2}} \, \frac{1}{ \Gamma - 50} \quad , \label{equ:GuettaetallikeGammadist}
\end{equation}
where the last factor comes from the assumption of a log normal distribution. In the cases FB-S and FB-D, the jet opening angle enters the probability distribution, which may be correlated with $\Gamma$. We use $n=1$ with the median value $\bar{x} = 2.6$ and the standard deviation $\sigma = 0.25$ to reproduce the observed distribution in \Ref~\cite{Guetta:2003wi}, which may be described, for instance, by $\theta \propto 1/\sqrt{\Gamma}$. Note that exactly this relationship has been shown to be consistent in \Ref~\cite{Ghirlanda:2011bn} with the Ghirlanda relation~\cite{2006A&A...450..471N} for the hypothesis FB-S. In the CB case, we use instead $n=2$ with $\bar{x} = 2.7$ and the standard deviation $\sigma = 0.25$, which can be theoretically motivated by the probability $\propto 1/\Gamma^2$ to observe a burst, coming from the forward collimation of the Lorentz cone.\footnote{Note that the offset of 50 is included to reduce the number of bursts with $\Gamma < 100$. }

In \figu{Gammadistr} we show the distributions for all of the three hypotheses. In each plot, the histogram shows the actual number of bursts generated with our Monte Carlo generator, while the thin dashed line is the theoretical distribution of the number of bursts depending on Lorentz factor as in \equ{GuettaetallikeGammadist}. The thick solid curves represent the relative contributions to the diffuse flux, which are the respective distribution functions times the appropriate scale factors for the models, as described above.  Even though all three histograms look similar by construction, the relative contributions exhibit significant differences. While the main contribution for the CB model comes from bursts with $\Gamma \approx 700 - 750$, the fireball models have their main contribution from lower values. The main contribution in the FB-S model comes from bursts with $\Gamma \approx 400$, which is still above the peak of the $\Gamma$ distribution. For the FB-D model it is even lower with $\Gamma \approx 140$ below the peak of the $\Gamma$ distribution. In none of the three cases the peak of the number distribution corresponds the peak in the relative contribution. Note that we have also checked that an intrinsic distribution of $\Gamma$, as it may be already expected within one object, such as a relativistically expanding fireball from which different components are observed with different viewing angles and therefore Doppler factors, is narrower than the distributions used here. In addition, note that
the contribution functions of CB and FB-S are qualitatively similar. For CB, the probability to observe a burst scales $\propto 1/\Gamma^2$ and the flux $E_\nu^2 \phi_\nu \propto \Gamma^4$, \ie, a total factor of $\Gamma^2$ survives in the contribution function. For FB-S, if $\theta$ is fixed, one has a flux $E_\nu^2 \phi_\nu \propto \Gamma^2$ with the same scaling.\footnote{In order to reproduce the empirical distribution, we effectively use $\theta \propto 1/\sqrt{\Gamma}$, which explains the difference in the contribution functions.} As a consequence, which can also be seen in \figu{Gammadistr}, for CB only few bursts contribute at the peak of the contribution function, whereas in the case FB-S many bursts contribute. Therefore, generically, one expects large fluctuations in stacked neutrino fluxes in the CB case, whereas in the fireball cases the fluctuations from $\Gamma$ are expected to be smaller. We will address this issue in \Sec~\ref{sec:lowstatistics}.

% FIGURE: RESULTS
\begin{figure}[t!]
\begin{center}
\includegraphics[width=\textwidth]{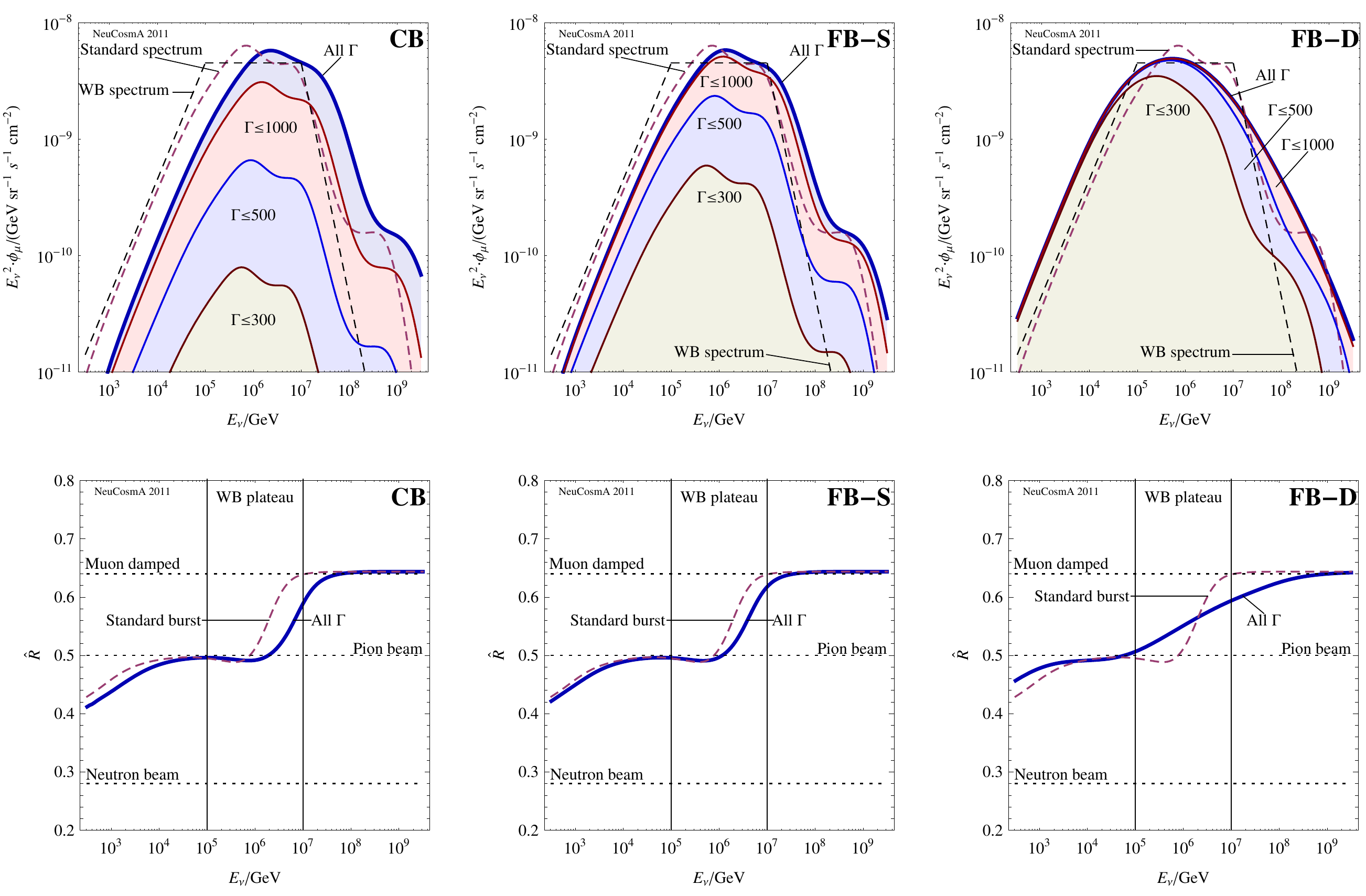}
\end{center}
\mycaption{\label{fig:Gammaresult} The diffuse fluxes of muon neutrinos  (upper panels) and the flavor ratios (lower panels) from $10 \, 000$ accumulated bursts. The upper plots show the contributions from all bursts with Lorentz factors $\Gamma \leq 300$, $\leq 500$, and $\leq 1000$ separately. The ``All $\Gamma$'' curves are normalized to the WB spectrum. The panels from left to right show the CB model, the FB-S model, and the FB-D model.  For comparison, we show the reference burst with values from \Tab~\ref{tab:params} as a dashed curve.}
\end{figure}

The resulting diffuse neutrino fluxes for the three hypotheses can be found in \figu{Gammaresult}. One can see from the left and the middle panels of \figu{Gammaresult} that the resulting fluxes and flavor ratios for CB (left panels) and FB-S (middle panels) are quite similar. Both show the distinct features of a single burst we already discussed in \Ref~\cite{Baerwald:2010fk}. The total fluxes (thick solid curves in upper panels) are just slightly shifted in energy because of the different scaling with Lorentz factor $\Gamma$. The effect of the different contribution functions is depicted by the thin solid curves representing the contributed flux from all bursts with $\Gamma \leq 300$, $\leq 500$, and $\leq 1000$, as labeled in the plots. The contribution from high $\Gamma$ bursts in the CB model is significantly higher than in the FB-S model because the peak contribution for CB comes from burst with $\Gamma \approx 700 - 750$, while for FB-S, it comes from bursts with $\Gamma \approx 400$.

For FB-D (right panels), the results are qualitatively very different.  This comes from the assumption of fixing all parameters at the detector, not in the SRF, as in the other two models. Thus, for each burst we have to calculate the isotropic volume with the corresponding collision radius and the corresponding shell thickness, see \eqs~\eqref{equ:ism} and \eqref{equ:visoISMexpl}. Moreover, all energies have to be boosted to the SRF with the correct Lorentz factor. 
In all burst simulations in the FB-D case, we scaled $C'_p$, $C'_{\gamma} \propto \Gamma^{-4}$, $\varepsilon'_{\gamma,\text{break}} \propto \Gamma^{-1}$, and $B' \propto \Gamma^{-3}$, \cf, \equ{BISM}. In consequence, the variation of the normalization factors leads to a dominance of low $\Gamma$ bursts with $\Gamma \approx 140$. Additionally, the change of the photon break energy in the SRF  $\varepsilon'_{\gamma,\text{break}}$ and the magnetic field $B'$ lead to a change of the flux shape. Thus, the features of a single burst are smeared out by summing over different bursts, and the flavor ratio transition broadens significantly. Note, however, that the smearing of magnetic field effects is the result of an implicitly assumed wide distribution of $B'$ with a similar effect as in the previous section (which was also estimated from the $\Gamma$ dependence of \equ{BISM}, applicable to FB-D).

In summary, the effects of a $\Gamma$ distribution are model-dependent. While most of the literature focuses on case FB-D, where magnetic field effects of the secondaries are smeared out, different results may be  expected even within the fireball model if the bursts are similar in the SRF. Since the observational data may not be mature enough to discriminate among the different cases, a distribution of $\Gamma$ has to be interpreted with care, and cannot be performed at a similar level as a distribution of redshift. On the other hand, it means  that some of the high-$\Gamma$ bursts accessible by \textit{Fermi}-LAT could indeed contribute more to the neutrino flux than previously anticipated. The test of magnetic field effects, such as the multi-peak structure of the spectral shape or a sharp transition of the flavor ratio, may help to discriminate FB-D versus CB and FB-S. In particular, the observation of these features may exclude conventional fireball phenomenology (FB-D).
However, CB and FB-S are intrinsically hardly distinguishable. Perhaps the statistical fluctuations of low statistics samples may give some clues, see discussion in \Sec~\ref{sec:lowstatistics}.

\subsection{Photon spectrum}

\begin{figure}[t!]
\begin{center}
\includegraphics[width=\textwidth]{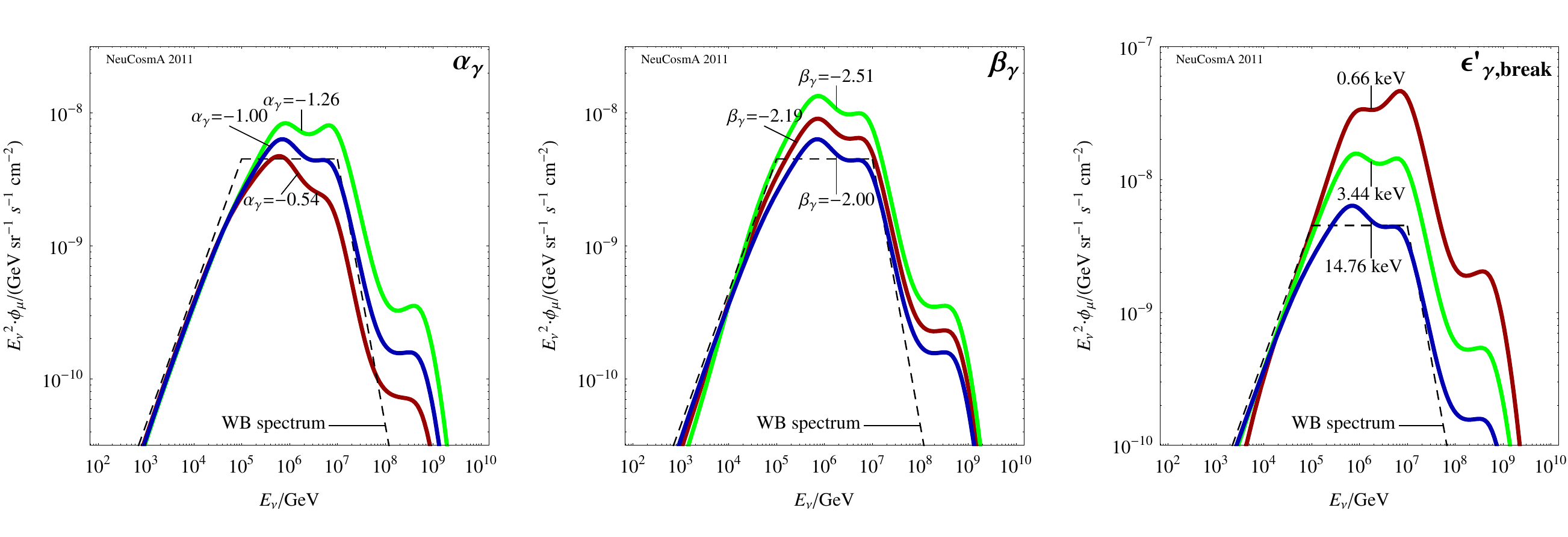}
\end{center}
\mycaption{\label{fig:basicdep} Neutrino flux for the reference values from \Tab~\ref{tab:params}, where the lower spectral index $\alpha_{\gamma}$ (left plot), the upper spectral index $\beta_{\gamma}$ (central plot), or the photon break energy $\varepsilon'_{\gamma,\text{break}}$ (right plot) of the target photon spectrum is varied. The spectra labeled $\alpha_{\gamma} = -1.00$, $\beta_{\gamma} = -2.00$, and $\varepsilon'_{\gamma,\text{break}} = 14.76 \, \kilo\electronvolt$, respectively, represent our standard burst as in \Tab~\ref{tab:params}, and are normalized to the WB spectrum. The other bursts are normalized by assuming that the total energy in photons is the same.}
\end{figure}

In the following, we test the effect of the photon spectrum on the neutrino flux, \ie, vary the parameters in \equ{targetphoton}. In \figu{basicdep}, the effect of $\alpha_{\gamma}$ (left plot), $\beta_{\gamma}$ (middle plot), and $\varepsilon'_{\gamma,\text{break}}$ (right plot) on the $\nu_\mu$ flux shape are shown. As before, the spectra with the standard parameters taken from \Tab~\ref{tab:params} are normalized by matching the integrated muon neutrino energy to the one in the WB flux. For the subsequent calculations we, however, fix the total energy in photons by integration over the whole spectrum. Before we come to the effect of the individual parameters, note that increasing the spectral indices or decreasing the break energy makes the spectra softer because of this normalization constraint. This means that more target photons for high energy protons are available, and the photohadronic interactions and therefore the neutrino production becomes enhanced. Although the photon spectrum may, in practice, be correlated with the other burst parameters in a non-trivial manner, this approach will turn out to be sufficient to demonstrate our main points. 

In the left and middle panels of \figu{basicdep}, we show the effect of $\alpha_\gamma$ and $\beta_\gamma$, respectively. As expected, the softer the spectrum is, the more neutrinos will be produced. For $\alpha_\gamma$, the neutrino spectral index above the first break varies, for $\beta_\gamma$, the neutrino spectral index below the first break, where these variations are anti-correlated with the photon spectral indices, as explained in \Sec~\ref{sec:qual}. For  $\varepsilon'_{\gamma,\text{break}}$ (right panel), a lower break energy in the photon spectrum leads to a higher $E_{\nu,\text{break}}$. The effect is, however, hardly visible on the left end of the spectra, because the muon and pion cooling breaks dominate the spectral shape, which are (in all cases) not shifted in energy by the photon spectrum. Only the relative contributions between muon peak and pion peak may change with $\varepsilon'_{\gamma,\text{break}}$, because the relative importance of the multi-pion processes is altered.

\begin{figure}[t!]
\begin{center}
\includegraphics[width=\textwidth]{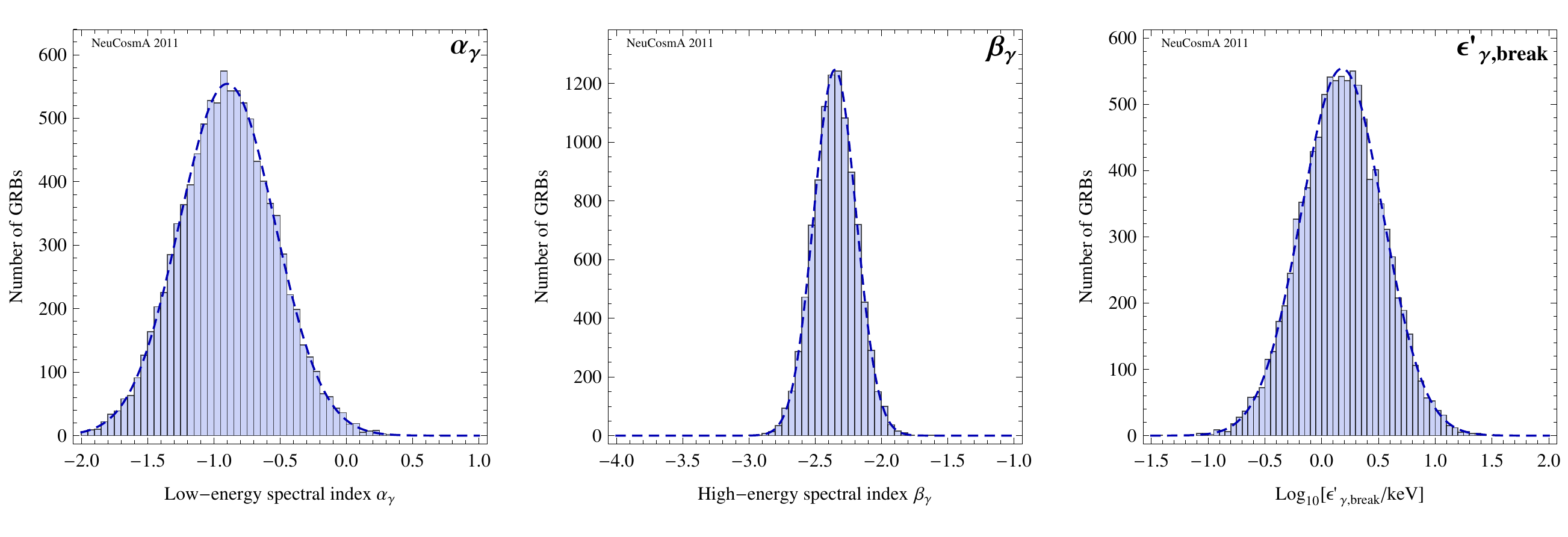}
\end{center}
\mycaption{\label{fig:specparamdistr} The distributions of the spectral parameters, namely the first spectral index $\alpha_{\gamma}$ (left plot), the second spectral index $\beta_{\gamma}$ (middle plot), and the photon break energy $\varepsilon'_{\gamma,\text{break}}$ (right plot). The histograms depict the $10 \, 000$ randomly picked values following the distributions obtained in \Ref~\cite{Nava:2010ig}. The dashed curves represent the analytic version of the distribution function.}
\end{figure}

We now want to introduce the distribution of the different parameters. Compared to the often used standard values $\alpha_\gamma=-1$ and $\beta_\gamma=-2$, we use data of long duration GRB obtained in \Ref~\cite{Nava:2010ig}, based on \textit{Fermi}-GBM data, to model our distributions, which are shown in \figu{specparamdistr}. The distribution for $\alpha_\gamma$  peaks at $\alpha_{\gamma} = -0.90$, is Gaussian, and has a standard deviation of $\sigma_{\alpha} = 0.36$. Therefore, our standard value is still in the $1\sigma$ range of the measured peak from that catalogue. For $\beta_\gamma$, there are only 84 long bursts in the data catalogue with a time-integrated spectrum best fit with the Band function. We have fitted a Gaussian to these values with a mean value $\beta_{\gamma} = -2.35$ and $\sigma_{\beta} = 0.16$. Here, our standard value is only within the $3\sigma$ range of the distribution. However, the statistics are considerably lower compared to the amount of data used to obtain the distributions for $\alpha_{\gamma}$ and $\varepsilon'_{\gamma,\text{break}}$. For $\varepsilon'_{\gamma,\text{break}}$ in $\kilo\electronvolt$, the peak of the distribution is on a logarithmic scale at $0.18$ ($10^{0.18} \, \kilo\electronvolt = 1.51 \, \kilo\electronvolt$) with a standard deviation $\sigma = 0.36$. As one can see the observed break energies are significantly lower than our standard value of $\varepsilon'_{\gamma,\text{break}} = 14.76 \, \kilo\electronvolt$ we obtained by assuming $E_{\nu,\text{break}} = 10^5 \, \giga\electronvolt$. Thus, the observed bursts should have higher values of $E_{\nu,\text{break}}$.\footnote{Note that in \Ref~\cite{Nava:2010ig} the distribution for the observed $E_{\text{peak}}$ is given, and not for the break energy in the SRF, $\varepsilon'_{\gamma,\text{break}}$. Therefore, we must boost the distribution from \Ref~\cite{Nava:2010ig} to the shock rest frame under consideration of our standard values for the Lorentz factor $\Gamma$ and the redshift $z$, see \Tab~\ref{tab:params}. Here, we neglect the slight difference between $E_\text{peak}$ from the Band function parameterization~\cite{Band:1993eg} and $\varepsilon_{\gamma,\text{break}}$ of the broken powerlaw parameterization in \equ{targetphoton}.}

\begin{figure}[t!]
\begin{center}
\includegraphics[width=\textwidth]{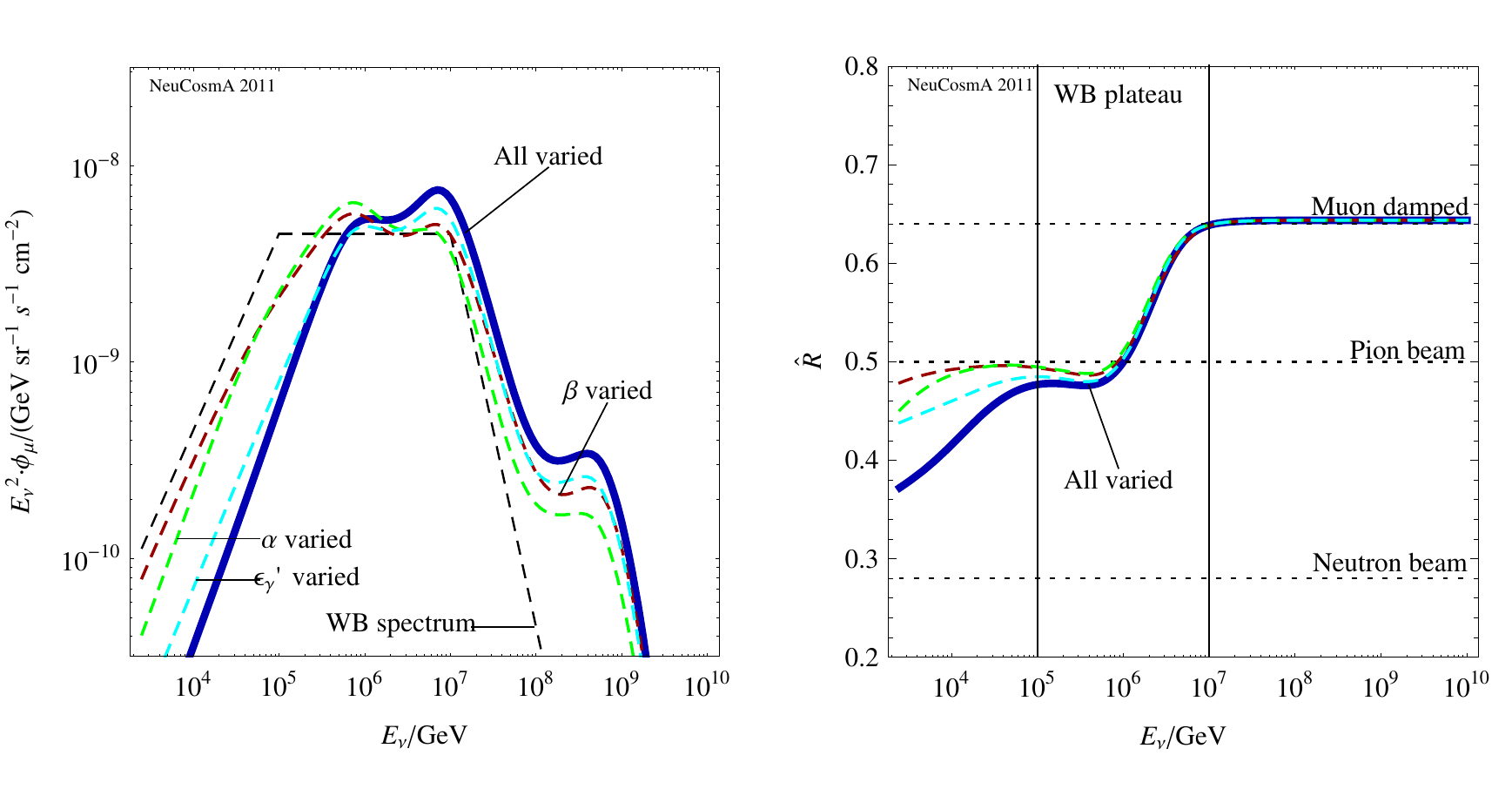}
\end{center}
\mycaption{\label{fig:photonspectrumresult} Neutrino flux (left plot) and flavor ratio (right plot) for a diffuse muon neutrino flux obtained by summing over $10 \, 000$ individual bursts with varied parameters of the input photon spectra. For the thick solid curve, denoted as ``All varied'', both photon indices, $\alpha_{\gamma}$ and $\beta_{\gamma}$, as well as the break energy $\varepsilon'_{\gamma,\text{break}}$ are varied in every spectrum. However, the total energy in photons is considered constant and each spectrum is normalized accordingly (by integration). In the left plot, the thin solid curves represent the resulting (quasi-)diffuse flux if only $\alpha_{\gamma}$, $\beta_{\gamma}$, or $\varepsilon'_{\gamma}$ is varied. The curves are labeled accordingly. The thin dashed curve represents the WB spectrum, the dashed curve the standard spectrum we obtained earlier.}
\end{figure}

The resulting neutrino fluxes and flavor ratios from the distributions in \figu{specparamdistr} are shown in \figu{photonspectrumresult} in the left and right panels, respectively. Here, the parameters  $\alpha_{\gamma}$, $\beta_{\gamma}$, and $\varepsilon'_{\gamma,\text{break}}$ have been varied separately (thin dashed curves) and simultaneously (thick solid curves). As the main result, the flavor ratio is hardly affected by the photon spectrum variation, and the generic structure of the flux shape also does not change. The most significant change may come from the different mean value of $\varepsilon'_{\gamma,\text{break}}$ compared to our reference burst, and the relative enhancement of the second peak coming from pion decays. These parameters, however, do not have any impact on our line of argumentation. The reason is that the peaks in the SRF are determined by \equ{ec}, which is completely independent of the shape of the target photon spectrum.

\section{Quasi-diffuse fluxes from low statistics samples} 
\label{sec:lowstatistics}

Here, we discuss the aggregation over a finite number of bursts, as it is used in a stacking analysis, see, \eg, \Ref~\cite{Abbasi:2011qc}. We use the same method as before, but we vary the discrete number of bursts $n$ used to compute a quasi-diffuse flux. Obviously, the statistical fluctuations coming from the summation over significantly fewer bursts than in a diffuse flux will introduce a systematical error in the extrapolation from the burst sample to a quasi-diffuse flux. However, it is exactly this extrapolation which is needed to compare the limits from neutrino telescopes to theoretical predictions, such as the Waxman-Bahcall estimate~\cite{Waxman:1998yy}, and to compare to predictions from fireball phenomenology. In this section, we quantify this systematical error using the redshift distribution, since we have shown that conclusions based on redshift may be drawn in the least model-dependent way.  However, we also demonstrate how statistics can be used to discriminate among different hypotheses for $\Gamma$. 

\begin{figure}[t!]
	\centering

	\includegraphics[width=\textwidth]{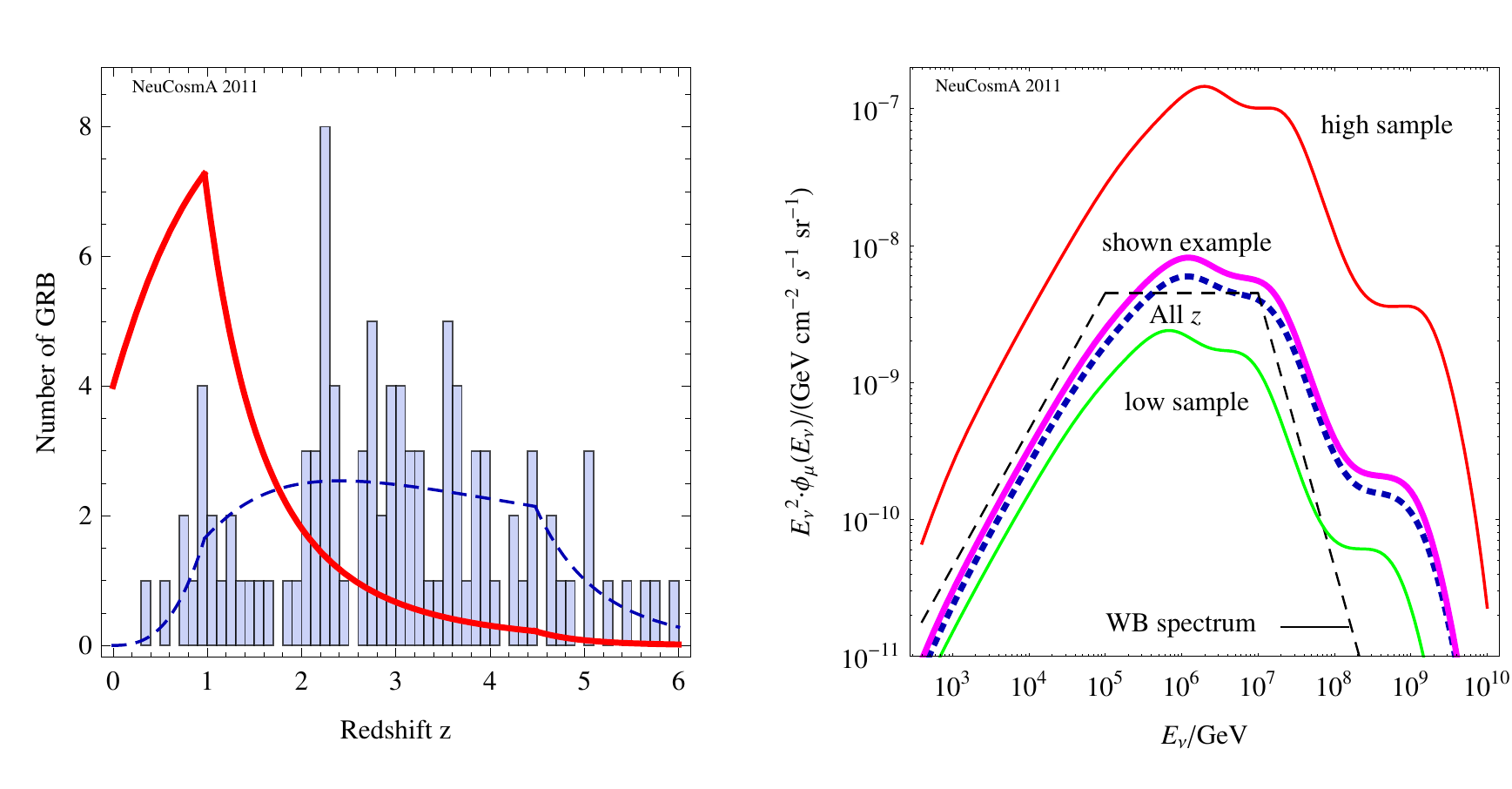}
	\mycaption{\label{fig:lowstatzsample}Example for a smaller subset of bursts. In the left panel, one sample of 100 randomly picked bursts for the strong evolution case is shown (histogram), for which the distribution function is illustrated as dashed curve. The solid curve is the (theoretical) contribution of a redshift value to the diffuse flux (scale arbitrary).  In the right panel, the thick solid curve shows the resulting quasi-diffuse flux for this sample. This has to be compared to the thick dashed curve, which refers to the high statistics result for the variation in redshift, see \figu{redshiftresult}. The other (thin) solid curves show the result for other subsamples, where the curves labeled ``high sample'' and ``low sample'' show extreme possibilities obtained in about 0.1\%  of all cases only. Moreover, the classic WB flux shape is shown as a thin black dashed curve.} 

\end{figure}

In \figu{lowstatzsample} we show one example for a sample of 100 randomly picked bursts (left panel). From the comparison between the histogram and the contribution function (solid curve), which depicts the weighted contribution of individual bursts, it is clear that strong fluctuations can be expected. The resulting quasi-diffuse flux for this distribution is shown in the right panel (thick solid curve), which somewhat deviates from the diffuse high statistics limit (dashed curve) because of these statistical effects. In addition, two extreme cases of samples of 100 bursts are shown as thin solid curves; such extreme cases are only obtained in about 0.1\% of all cases. If other (independent) parameter distributions are present, such as for $B'$ or $\Gamma$, we expect that the resulting statistical error, to a first approximation, adds up in a Gaussian manner. Therefore, the redshift distribution will only give a lower estimate for the introduced systematical error. A more detailed discussion including the impact of the gamma-ray detection threshold on a stacking analysis using a simultaneous redshift-luminosity variation can be found in \App~\ref{sec:threshold}.

\begin{figure}[t!]
\centering
\includegraphics[width=0.49\textwidth]{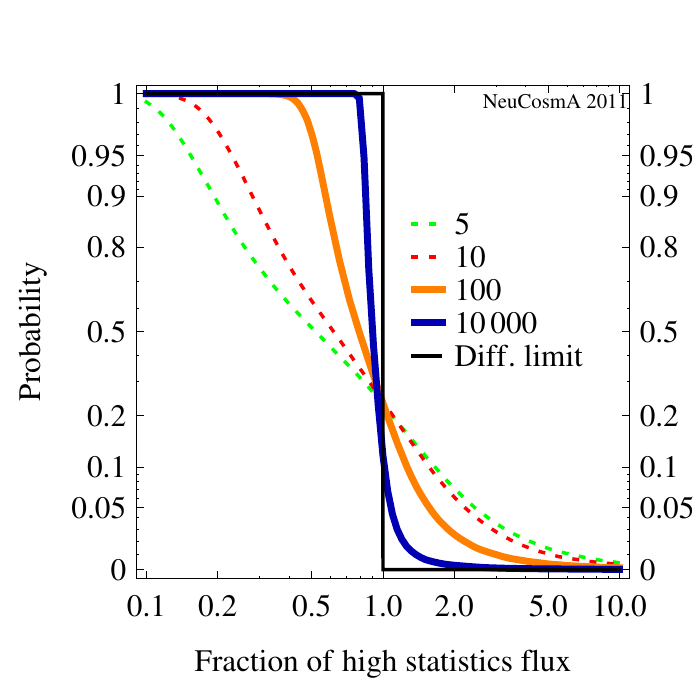}
\mycaption{\label{fig:lowstatz} Probability that the quasi-diffuse flux extrapolated from a low statistics sample with $n$ bursts is larger than a certain fraction of the diffuse limit (see legend for different values of $n$). This function corresponds to (one minus) the cumulative distribution function of the probability density. Here, the strong evolution case for the redshift is used for the burst distribution function.}
\end{figure}

While this figure illustrates that some effect of the statistical fluctuations can be expected, which depend on the sample size, it is not sufficient for a quantification of the problem. Therefore, we choose the following method: We show in \figu{lowstatz} the probability that the resulting quasi-diffuse flux from a low statistics sample with $n$ bursts is larger than a certain fraction $f$ of the diffuse limit.\footnote{Methodologically, we first systematically generate bursts with $0 \leq z \leq 6$ in steps $\Delta z = 0.001$ and calculate the energy flux density. Then, in the second step, we draw samples of $n$ bursts with probabilities according to \equ{Burstrateredshift} (with strong evolution). In order to compute the probabilities in the figure, we repeat this procedure for each sample size $N$ times, where, ideally, $N \rightarrow \infty$, and, practically, large enough $N = 100 \, 000$ are used. The used energy flux density $\int E_{\nu} \, \phi_{\mu}(E_{\nu}) \, \mathrm{d}E_{\nu}$ is obtained by integration performed from $10^0 \, \giga\electronvolt$ to $10^{10} \, \giga\electronvolt$.}
 By definition, the probability that the diffuse flux is larger than the fraction $f<1.0$ of the diffuse limit is always one, whereas it is zero for $f>1.0$, which is shown as the step function labeled ``diff. limit'' in the figure. The smaller the sample size is, the stronger is the deviation from this curve. For example, for $n=5$, the probability that $f \le 0.2$, \ie, at most 20\% of the diffuse limit are obtained as a result, is about $1-0.89=0.11$, and the probability that $f \ge 5.0$, \ie, five times more than the diffuse limit is obtained, is about 0.01. While this figure is somewhat difficult to read, it contains already all of the information discussed in the following. Note that this method corresponds to computing one minus the cumulative distribution function of the probability density.
 
\begin{table}[t]
  \centering

\includegraphics[width=\textwidth]{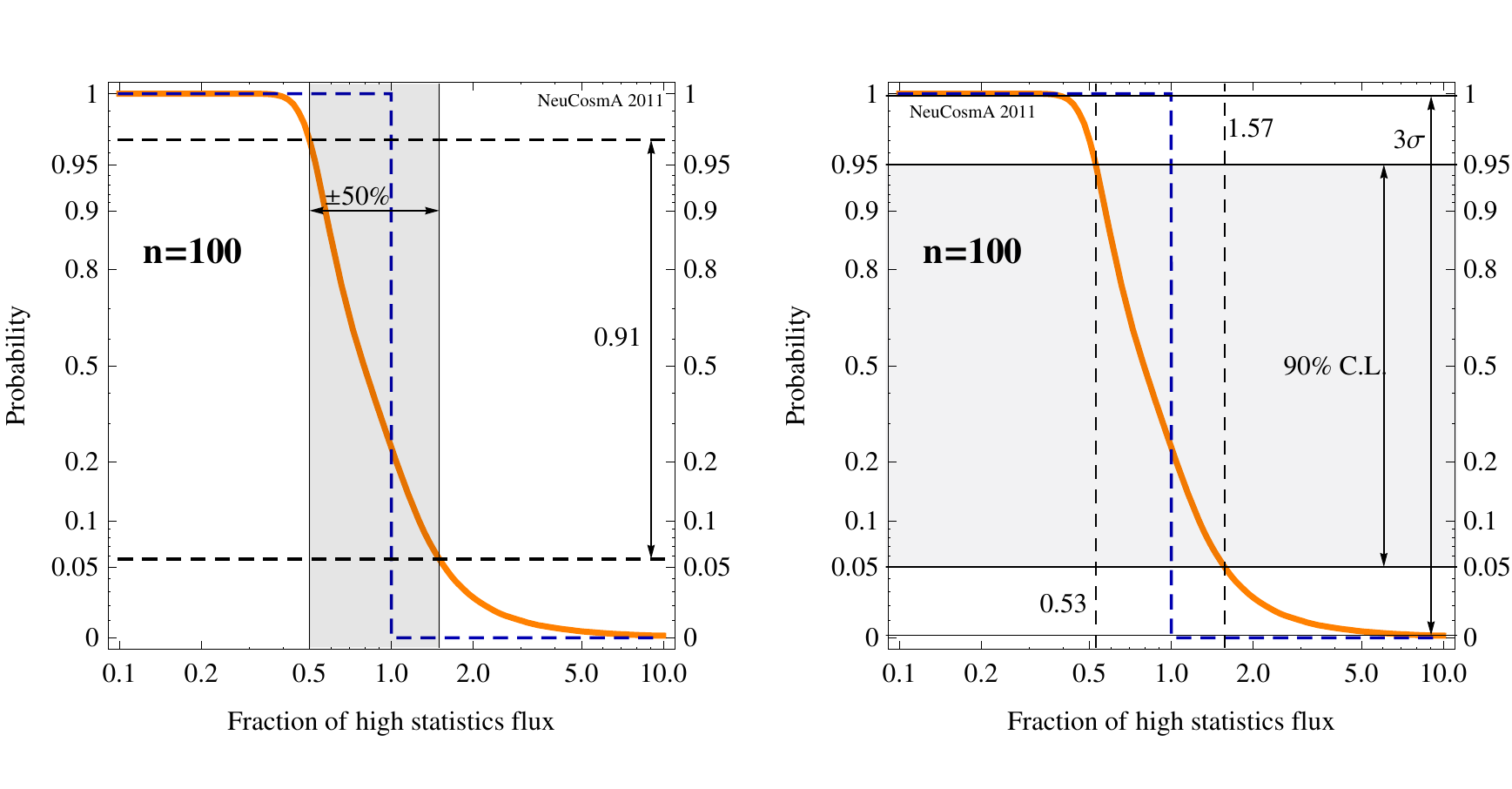}

\vspace*{0.5cm}
  \begin{tabular}{r|ccc}
    \hline
    $n$ & $\pm 10 \%$ & $\pm 20\%$ & $\pm 50\%$ \\%& $\pm 10 \%$ & $\pm 50\%$ & $\pm 10 \%$ & $\pm 50\%$ \\
    \hline
  
    5 & $0.07$ & $0.14$ & $0.40$ \\%& $0.04$ & $0.25$ & $0.14$ & $0.71$ \\
    10 & $0.09$ & $0.18$& $0.52$ \\%& $0.06$ & $0.34$ & $0.19$ & $0.85$ \\
    50 & $0.14$ & $0.30$ & $0.82$ \\%& $0.11$ & $0.61$ & $0.38$ & $0.99$ \\
    100 & $0.17$ & $0.37$ & $0.91$ \\%& $0.14$ & $0.75$ & $0.51$ & $1.00$ \\
    300 & $0.23$ & $0.50$ & $0.96$ \\%& $0.20$ & $0.91$ & $0.73$ & $1.00$ \\
    1000 & $0.30$ & $0.69$ & $0.98$ \\%& $0.29$ & $0.95$ & $0.90$ & $1.00$ \\
    10000 & $0.48$ & $0.97$ & $0.99$ \\%& $0.62$ & $1.00$ & $1.00$ & $1.00$ \\
  \hline
  \end{tabular}
\hspace*{1cm}
\begin{tabular}{r|cc}
    \hline
    $n$ & Rel. error $90 \%$ CL & Rel. error $3\sigma$ \\
    \hline
    5 & $0.15 - 2.45$ & $0.09 - 20.89$ \\
    10 & $0.23 - 2.22$ & $0.13 - 19.05$ \\
    50 & $0.44 - 1.72$ & $0.30 - 10.28$ \\
    100 & $0.53 - 1.57$ & $0.39 - 8.78$ \\
    300 & $0.64 - 1.38$ & $0.53 - 6.53$ \\
    1000 & $0.72 - 1.25$ & $0.64 - 5.15$ \\
    10000 & $0.83 - 1.08$ & $0.78 - 2.62$ \\
    \hline
  \end{tabular}
  \mycaption{\label{tab:lowstatprobabilities} Lower row, left table: Probability that the quasi-diffuse flux is within a certain interval of the diffuse limit for different stacking sample sizes $n$. Right table: Systematical error coming from the extrapolation of the quasi-diffuse flux from a sample with $n$ bursts at different confidence levels.
The corresponding figures (upper row) illustrate how this information is obtained from \figu{lowstatzsample} for $n=100$.
 Here, the strong evolution case for the redshift is used for the burst distribution function. 
}
\end{table}

For example, one can ask what the probability for a quasi-diffuse flux, derived from the stacking of $n$ bursts, is to be within a certain fraction of the diffuse (high statistics) limit. This question is addressed in the left column of \Tab~\ref{tab:lowstatprobabilities}.  The plot in the upper left panel illustrates how this information is obtained from \figu{lowstatzsample}: for a certain interval around the diffuse limit curve, the corresponding probability contained in this interval is read off from the vertical axis. From the table, one finds that $n=10 \, 000$ is already close to the diffuse limit. Quantitatively, the probability that a quasi-diffuse flux based on $10 \, 000$ bursts, such as based on the number of observable bursts during ten years of IceCube operation, is within 20\% of the diffuse limit from a purely statistical point of view, is 97\%.  That is sufficiently diffuse for any practical purpose, but, nevertheless, it is surprising that small fluctuations are expected. Practically, it means that the diffuse flux during ten years of IceCube operation may not be exactly the same as the one during the following ten years at the level of 20\%. One can also read off from that table that for $n=100$, as used in recent IceCube stacking limits, the probability to be within 20\% of the diffuse limit is only 37\%. Only if $n = \mathcal{O}(1000)$, reliable predictions from stacking analyses can be expected. 

A different but complementary question is addressed in the right column of \Tab~\ref{tab:lowstatprobabilities}. Here, the variation of the flux with a certain probability window is discussed, see upper right figure for an illustration. Since the curve shown in this figure corresponds to the integrated probability density, it can be used to derive the range (horizontal axis) in which 90\% of all cases and 99.73\% of all cases are found. These numbers can be used to estimate the systematical error in the diffuse-flux extrapolation at the corresponding confidence level. The result is shown in the lower right panel in \Tab~\ref{tab:lowstatprobabilities}. It can be inferred from this table that the systematical error in the calculation of the quasi-diffuse flux in terms of the energy flux density (and thus, at least roughly, in terms of the flux normalization) is at the level of about 50\% for 100 bursts, 35\% for 300 bursts, and 25\% for 1000 bursts (90\% CL) from the redshift distribution only -- additional parameter variations will lead to a larger effect (see \App~\ref{sec:threshold}). This is to be compared with the naive statistical estimate if all bursts were at the same $z$: the $1 \sigma$ (90\% CL) relative error for 100 bursts may be estimated as $1/\sqrt{100} \simeq 0.1$ (0.16), to be compared to the actually obtained 50\% at the 90\% CL.

Apart from the discussion of finite sample sizes, the special case $n=1$ is interesting. In particular, what is the probability that a single bright burst is detected in ten years of IceCube operation, given the redshift distribution (strong evolution case) only? In this case, the multi-messenger connection to gamma-rays may not be necessary, and the burst should also outshine the backgrounds. As a prerequisite, a few events from such a single burst, say at least three, have to be detected. The probability for such a burst depends on the level of the diffuse flux. If it saturates the WB estimate, such a burst would need to contribute only about $1/30$ of the total flux.\footnote{This can be derived from the estimate that the WB neutrino flux, if saturated, leads to $\mathcal{O}(100)$ events in ten years of operation of IC-86, which is consistent with the current IC-40 bound.} If the diffuse flux is only at the level of $1/10$ of the WB estimate, such a burst has to contribute $1/3$ of the total flux. From these numbers, we can compute the maximally allowed redshift, which is $z_{\text{max},1} \simeq 0.14$ and $z_{\text{max},2} \simeq 0.05$, respectively, for these two different cases. As the next step, we obtain as probabilities for a single burst $P(z \leq 0.14) = 5 \cdot 10^{-5}$ and $P(z \leq 0.05) = 2 \cdot 10^{-6}$. Given that $\mathcal{O}(10 \, 000)$ bursts are expected in the observable universe in ten years, the probability that such a bright burst, which does not require the observation of the gamma-ray counterpart, will occur at least once is $1-[1-P(z \le z_{\text{max}})]^n$ with $n=10 \, 000$, which is about 40\% if the WB flux is saturated, and about 2\% if the diffuse neutrino flux is an order of magnitude smaller.

\begin{figure}[t!]
\centering
\includegraphics[width=0.49\textwidth]{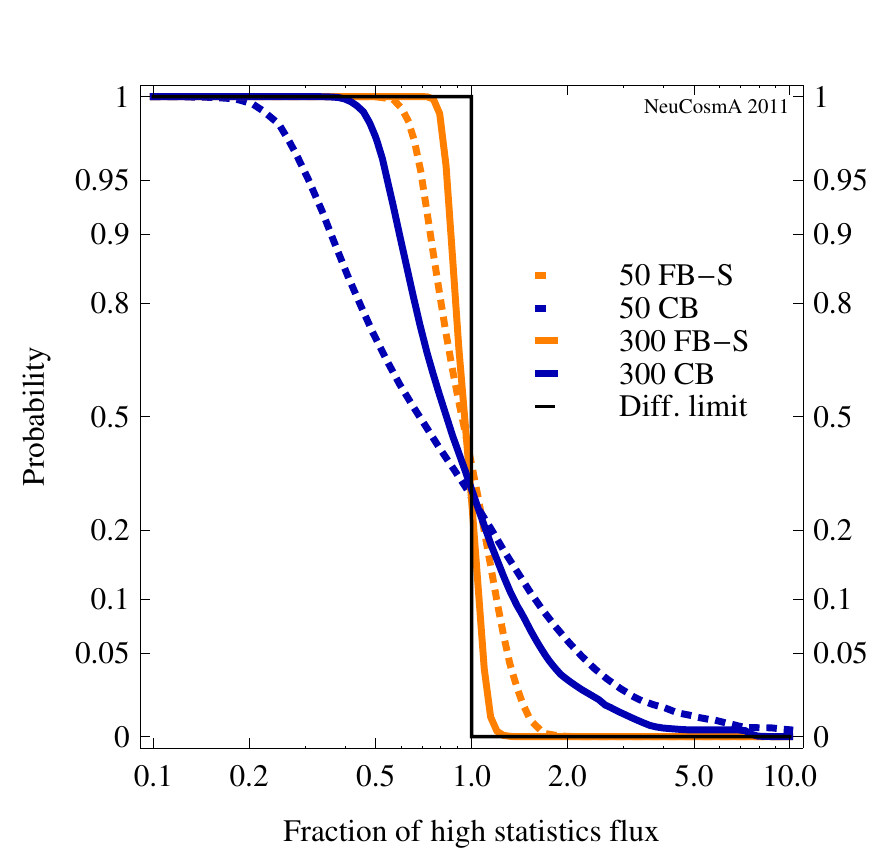}
\mycaption{\label{fig:lowstatGamma} Figure similar to \figu{lowstatz}, but for a variation of the Lorentz factor $\Gamma$ for the cases CB and FB-S.  Here, the two models are compared for $n=50$ and $n=300$.}
\end{figure}

In \Sec~\ref{sec:lorentz}, we have discussed that the consequences of variations of $\Gamma$ are model-dependent. However, we have illustrated that the models CB and FB-S lead to relatively similar fluxes in the quasi-diffuse limit, whereas for low statistics, stronger fluctuations are generically to be expected for CB than for FB-S (\cf, \figu{Gammadistr}, where the bursts statistics close to the peaks of the distribution functions are different). We test this hypothesis quantitatively in \figu{lowstatGamma}, where we directly compare the integrated probability distributions between CB and FB-S for two sample sizes ($n=50$ and $n=300$). It can be clearly seen that the case CB deviates much stronger from the diffuse limit (step function), as expected. In addition, this difference between the two models becomes smaller for larger $n$, \ie, in the high statistics limit, there is hardly any difference.  For example, for $n=50$, the probability to be within $\pm 10\%$ of the diffuse limit is only 10\% for CB, whereas it is about 40\% for FB-S. This statistical effect suggests to use this effect for a test of the model: For example, one may work with randomly chosen $n=50$ samples from a larger stacking sample accumulated over many years. Then by the test of the variance of the neutrino fluxes among these samples one may draw conclusions on the underlying model. Note that CB/FB-S versus FB-D can be discriminated by the impact on the magnetic field effects, as we have discussed earlier.

\section{Summary and conclusions}

We have numerically computed GRB neutrino fluxes and flavor ratios including different meson production modes, magnetic field effects, and flavor mixing. Our starting point has been the minimal set of assumptions needed to compute the neutrino fluxes at the same level as it is used in recent IceCube stacked limits, but including the full spectral dependencies and the cooling of the secondaries explicitely. We have kept these assumptions as model-independent as possible. As the next step, we have connected the quantities needed for the neutrino production to different models, such as a cannonball-like model and conventional fireball phenomenology, where we have clearly demonstrated where model-dependent assumptions or relationships enter. 

The main purpose of this study has been the computation of aggregated fluxes, where we have distinguished diffuse fluxes and quasi-diffuse fluxes obtained from stacking analyses. For the computation of the aggregated fluxes, we have computed the flux burst by burst, where we have used $10 \, 000$ bursts for the diffuse flux -- as may be expected in the observable universe within ten years of neutrino telescope operation. We have established that these $10 \, 000$ bursts lead to a flux close enough to the diffuse limit, \ie, that the systematical error coming from the burst statistics is small in that case. Quantitatively, from the redshift distribution (strong evolution case), the neutrino flux in these ten years will be within 20\% of the diffuse limit with 97\% probability.

Our computations of aggregated fluxes have been based on parameter distributions for redshift $z$, magnetic field $B'$, Lorentz boost $\Gamma$, and the spectral shape of the target photons. We have identified two major issues: The theoretical distribution of the parameter, and the interpretation of a different parameter value in terms of the relative contribution to the flux. A relatively simple case is redshift, which may follow the star formation rate in some form, and the interpretation of the relative contributions of the bursts, if assumed to be otherwise comparable at the source, is straightforward. A completely different case is $\Gamma$ for which the theoretical distribution is unknown and the translation of the relative contribution of the bursts is model-dependent.
For example, in the conventional fireball model, the  size of the interaction region is estimated from the observed variability timescale $t_v$, which implies that it strongly depends on the Lorentz factor $\Gamma$. As a consequence, small values of $\Gamma$ are assumed to dominate the neutrino flux. In the same model, the hypothesis of similar properties in the comoving frame leads to the dominance of high $\Gamma$, as it is consistent with the recent Ghirlanda et al. study~\cite{Ghirlanda:2011bn}. As a consequence, some of the \textit{Fermi}-LAT accessible bursts with high $\Gamma$ may contribute stronger to the neutrino flux than previously anticipated.
Since the underlying assumptions for the different parameter distributions are qualitatively different, we have not mixed these different cases in the computation of the diffuse fluxes; in addition, our quantitative statements (see below) are based on the redshift distribution only, as the most conservative assumption.

The synchrotron cooling versus decay of the secondaries (pions, muons, kaons) in combination with multi-pion photohadronic processes leads to a characteristic multi-peak structure of the GRB neutrino flux. In addition, a transition  of the flavor composition from a pion beam source (neutrinos from pion decay chain) to a muon damped source (neutrinos from muon decays suppressed) at high energies is expected, which can, in principle, be measured by the muon track to cascade ratio in a neutrino telescope. We have established that these features are, in general, also present in diffuse fluxes, such as if the redshift of the bursts is varied. Only too strong variations of the magnetic field may destroy the spectral peaks in the flux and smear out the flavor ratio transition. In this case, the spectral shape of the flux cannot be used anymore to obtain information on the magnetic field, whereas the flavor composition depends on the mean value of $B'$ and the broadness of its distribution with no other significant parameter dependencies. This means that the flavor ratio can be regarded as the most robust prediction one can make for diffuse GRB neutrino fluxes. The observation of magnetic field effects can also be useful as a model discriminator: we have demonstrated that the observation of spectral features in diffuse GRB neutrino fluxes or a narrow transition of the flavor ratio can discriminate between the conventional fireball assumption that bursts with low $\Gamma$ dominate the neutrino flux and the hypothesis that the bursts are alike in the comoving frame.

Apart from diffuse fluxes, we have also discussed quasi-diffuse limits obtained from the stacking of a relatively  small number of bursts, such as $\mathcal{O}(100)$ bursts used in recent IceCube analyses. The stacking statistics of a small number of bursts has been shown to introduce a systematical error in the computation of the quasi-diffuse flux, which is at the level of about 50\% for 100 bursts, 35\% for 300 bursts, and 25\% for 1000 bursts (90\% CL) from the redshift distribution only -- additional parameter variations will lead to a larger effect (see \App~\ref{sec:threshold}). Therefore, $\mathcal{O}(1000)$ bursts may have to be stacked for reliable conclusions.
The reason is that the redshift $z$ leading to the largest contribution to the flux does not coincide with the maximum of the redshift distribution. As another consequence, in stacking analyses, it may also be consistent to assign $z \simeq 1$ to a redshift not measured for a burst, since, depending on the method, a different value may lead to an overestimate of the neutrino flux or to wrong energies of the spectral neutrino breaks. This conclusion holds almost independently of the redshift distribution, including the strong evolution case. Similar standard values can be inferred for $\Gamma$, which are, however, model dependent. For example, $\Gamma \simeq 700$ gives the main contribution in our cannonball ansatz, $\Gamma \simeq 400$ in the fireball model with the parameters fixed in the comoving frame, and  $\Gamma \simeq 140$ for the conventional fireball assumption. While normally a high stacking statistics is desirable, we have also demonstrated that indeed the fluctuations among different stacking samples may be used as a discriminator between cannonball and fireball approaches.

A very interesting special case is the observation of a single bright burst detectable without multi-messenger counterpart. From the redshift distribution (strong evolution case), we have estimated that the probability to observe at least one single bright burst individually detectable in IceCube during ten years of full-scale operation is about 40\% if the Waxman-Bahcall flux is saturated, and 2\% if the diffuse neutrino flux is an order of magnitude smaller. Thus, at best, there may be a 50-50 chance for such a single burst detection.

We conclude that neutrino observations of GRBs, both in terms of flux and flavor ratio, may not only help to test the paradigm if GRBs are the sources of the highest energetic cosmic rays, but also give important clues on the underlying models. Especially spectral features and the flavor composition turn out to be very interesting discriminators of very basic assumptions if a signal is observed. If no signal is observed, a systematical error coming from the stacking statistics has to be regarded in the context of the limit calculation.

\subsubsection*{Acknowledgments}

We would like to thank M. Ahlers, J. K. Becker, A. Kappes, P. Mehta, K. Murase, and E. Waxman for useful and illuminating discussions. 
 PB and SH acknowledge support  from the Research Training Group GRK1147 ``Theoretical astrophysics and particle physics'' of Deutsche Forschungsgemeinschaft, SH from the Studienstiftung des deutschen Volkes (German National Academic Foundation), and WW from the Emmy Noether program of Deutsche Forschungsgemeinschaft, contract WI 2639/2-1.

%\bibliographystyle{h-elsevier}
%\bibliography{references}

\newpage 

\appendix

\section{Details of the source model}
\label{sec:details}

In this appendix, we describe details of the source model.

\subsection{Components of the model}
\label{sec:components}

In the following, primed quantities refer to the shock/bulk rest frame (SRF), and unprimed quantities to the observer's frame. In some cases, which we denote explicitely, the unprimed quantities may also refer to the source (engine) frame.

\subsubsection*{Photon and proton input spectra}

We assume that the target proton field in the SRF is given by \equ{targetproton}.
Here, the maximal proton energy  $E'_{p,\text{max}}$ is derived from the comparison between acceleration rate of the protons~\cite{Hillas:1985is} 
\begin{equation}
	t'^{-1}_{\text{acc}} = \eta \frac{c^2 \; q \; B'}{E'} \, ,\label{eq_acctime}
\end{equation}
and the synchrotron loss rate
\begin{equation}
	t'^{-1}_{\text{syn}}(E') = \frac{e^4 \; B'^{2} \; E'}{9\pi \; \varepsilon_0 \; m^4 c^5} \, , \label{equ:synchrotrontime}
\end{equation}
leading to
\begin{equation}
	E'_{p,\text{max}} 
	 \simeq 2 \cdot 10^{8} \, \left( \frac{\eta}{0.1} \right)^{\frac{1}{2}} \, \left( \frac{B'}{10^5 \, \text{G}} \right)^{-\frac{1}{2}} \, \giga\electronvolt \quad , \label{equ:epmax}
\end{equation}
where $\eta$ is an acceleration efficiency.
This assumption is somewhat ``vanilla'' in the presence of adiabatic cooling (or the dominance of photohadronic cooling), since the maximal proton energy may be limited otherwise. However, since the neutrino spectrum will be dominated by the energy losses of the pions for $E_\nu \gtrsim 10^7 \, \mathrm{GeV}$, we expect these effects to be small.

The target photon field $N'_{\gamma}(\varepsilon')$ is assumed to be derived from observed GRB photon spectra and is parameterized by \equ{targetphoton}. The spectral indices $\alpha_{\gamma}$ and $\beta_{\gamma}$ can be obtained from GRB observations.  The break energy $\varepsilon'_{\gamma,\text{break}}$ is normally retrieved from the photon spectrum of a burst, but we will need it as a parameter in our model. Moreover, the energy $\varepsilon'_{\gamma,\text{max}}$ is the maximum photon energy, and $\varepsilon'_{\gamma,\text{min}}$ is the lower energy cut-off of the photon spectrum. 
In practical applications, these may be chosen according to the energy window of the gamma-ray observation. Here, we choose $\varepsilon'_{\gamma,\text{min}} = 0.2 \, \mathrm{eV}$, $\varepsilon'_{\gamma,\text{max}}=300 \,\varepsilon'_{\gamma,\text{break}}$~\cite{Lipari:2007su}, $E'_{p,\text{min}} \simeq 1 \, \mathrm{GeV}$, and $\eta=0.1$. These values are somewhat optimistic in the sense that the threshold for photohadronic interactions is always sufficiently passed in the whole energy range.  However, we have tested that the neutrino results are not very sensitive to $\varepsilon'_{\gamma,\text{min}}$ (as long as $\varepsilon'_{\gamma,\text{min}} \lesssim 1/30  \, \varepsilon'_{\gamma,\text{break}}$), to $\varepsilon'_{\gamma,\text{max}}$ (as long as $\varepsilon'_{\gamma,\text{max}} \gtrsim 30 \,\varepsilon'_{\gamma,\text{break}}$),  to $E'_{p,\text{min}}$ (as long as $E_{p,\text{min}} \lesssim 1000 \, \mathrm{GeV}$), and to $\eta$ (as long as $\eta \gtrsim 0.01$). Note that any violation of the conditions in brackets for the photon spectrum would be visible even within the BATSE energy window (given the Lorentz boost of these quantities).

\subsubsection*{Photohadronic interactions and weak decays}

The accelerated protons interact with the target photon field, where pions, kaons, and neutrons are produced. We consider the following processes:
\begin{align}
 p + \gamma & \rightarrow \pi + p' \, ,\\
 p + \gamma & \rightarrow K^+ + \Lambda/\Sigma \, .
\end{align}
Here, $p'$ is either a proton or neutron, $\pi$ are one or more neutral, positive, or negative charged pions, and $\Lambda$ and $\Sigma$ are resonances.
Note that the effect of kaon decays is usually small, but kaon decays may have interesting consequences for the neutrino flavor ratios at very high energies, in particular, if strong magnetic fields are present~\cite{Kachelriess:2006fi,Asano:2006zzb,Kachelriess:2007tr,Moharana:2011hh}.
Therefore, we consider the leading mode: $K^+$ production (for protons in the initial state) with the decay channel leading to highest energy neutrinos. Note that at even higher energies, other processes, such as charmed meson production, may contribute as well.

We follow the description of the photohadronic interactions in \Ref~\cite{Hummer:2010vx} (Sim-B), based on the physics of SOPHIA~\cite{Mucke:1999yb}. The treatment, taking into account resonant (including higher resonances), direct, and high energetic multi-pion production, allows for an efficient computation which is very accurate for power law-like spectra. It predicts well neutrino to antineutrino ratios and flavor ratios. The secondary meson injection rate $Q_b'(E_b')$ (in units of $\mathrm{GeV^{-1} \, cm^{-3} \, s^{-1}}$) for the interaction  between the proton spectrum $N_p'(E_p')$ in \equ{targetproton} and the target photon field $N_\gamma'(\varepsilon')$ in \equ{targetphoton} is given by
\begin{equation}
Q_b'(E_b') = \int\limits_{E_b'}^{\infty} \frac{dE_p'}{E_p'} \, N_p'(E_p') \, \int\limits_{\frac{\epsilon_{\mathrm{th}} m_p}{2 E_p'}}^{\infty} d\varepsilon \, N_\gamma'(\varepsilon') \,  R_b( x,y )  \, .
\label{equ:prodmaster}
\end{equation}
with $x=E_b'/E_p'$ the fraction of energy going into the secondary, $y \equiv (E_p'\varepsilon')/m_p$ (directly related to the center of mass energy), the ``response function'' $R_b( x,y )$, and $\epsilon_{\mathrm{th}}$ the threshold for the photohadronic interactions (photon energy in proton rest frame). Since many interaction types are considered, the response function can be quite complicated; for details, see \Ref~\cite{Hummer:2010vx}. 
From \equ{prodmaster}, one can make already two very important observations: First of all,  in $Q'_b(E_b')$ only the product of the normalizations of the proton and photon fluxes enters. This means that  the ratio between the energy in photons (or electrons) and protons, \ie, the baryonic loading, is not explicitely required for the overall normalization of the neutrino spectrum -- unless processes explicitely depending on the proton or photon density contribute significantly (\eg, for inverse Compton scattering). The second observation from \equ{prodmaster} is that the {\em steady} densities [$\giga\electronvolt^{-1} \, \centi\meter^{-3}$] of proton and photon fields in the interaction region determine the neutrino production, which we have used as initial input. These steady distributions are often (\eg, in \Ref~\cite{Guetta:2003wi})  directly computed from the ejected (observable) gamma-ray flux under the assumption $Q'_\gamma \simeq N'_\gamma/t'_{\mathrm{esc}}$  (see \equ{steadystateesc} below)  with $t'_{\mathrm{esc}}$ related to the size of the interaction region (\eg, shell width in the internal shock model). For the neutrino production, however, it does not matter if the photons escape the source or not, which is taken into account in some models such as the photospheric emission model, see, \eg, \Ref~\cite{Murase:2008sp}. For the model-independent part of our calculation, it does not matter.

We let the produced particle species, namely pions, kaons, and neutrons, decay weakly into neutrinos as
\begin{eqnarray}
\pi^+ & \rightarrow & \mu^+ + \nu_\mu \, ,\nonumber \\
& & \mu^+ \rightarrow e^+ + \nu_e + \bar{\nu}_\mu \, , \label{equ:piplusdec} \\
\pi^- & \rightarrow & \mu^- + \bar\nu_\mu \, , \nonumber \\
& & \mu^- \rightarrow e^- + \bar\nu_e + \nu_\mu \, , \label{equ:piminusdec} \\
 K^+ & \rightarrow & \mu^+ + \nu_\mu  \label{equ:kplusdec} \, ,\\
 n & \rightarrow & p + e^- + \bar{\nu}_e  \label{equ:ndec}  \, .
\end{eqnarray}
The weak decays are treated as described in \Ref~\cite{Lipari:2007su}, for details see also \Ref~\cite{Hummer:2010ai}. Since the decays of muons are helicity dependent, we distinguish between left and right handed muons, which has some impact on the flavor ratio.

\subsubsection*{Energy losses and escape}

\begin{figure}[t]
\begin{center}
 \includegraphics[width=0.5\textwidth]{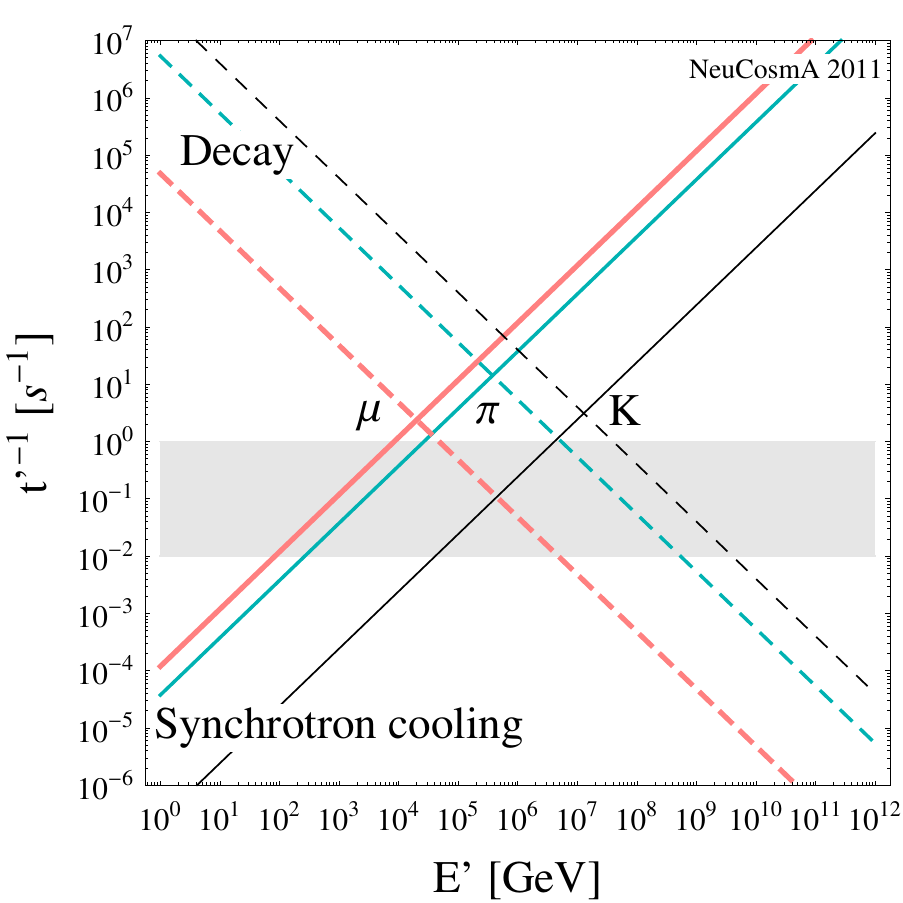}
\end{center}
 \mycaption{\label{fig:times} Synchrotron cooling and decay rates for pions, muons, and kaons as a function of energy for a magnetic field of $B' \simeq 300\,\mathrm{kG}$. The gray-shaded region shows the estimated range for the escape (or adiabatic cooling) rate if the size of the acceleration region is given by a shell width determined by a variability timescale ranging from $0.01 \, \mathrm{s}$ to $1 \, \mathrm{s}$ (Lorentz factor $\Gamma=300$, redshift $z=2$; \cf, \equ{ism}).}
\end{figure}

The kinetic equation for the particle spectrum, assuming continuous energy losses, is given by (see, \eg , \Ref~\cite{Atoyan:2002gu}):
\begin{equation}
\label{equ:steadstate}
Q'(E')=\frac{\partial}{\partial E'}\left(b'(E') \, N'(E')\right)+\frac{N'(E')}{t'_\mathrm{esc}} \, ,
\end{equation}
with $t'_\mathrm{esc}(E')$ the characteristic escape time, $b'(E')=-E' \, t'^{-1}_{\mathrm{loss}}$ with
$t'^{-1}_{\mathrm{loss}}(E')=-1/E' \, dE'/dt'$ the rate characterizing energy losses, $Q'(E')$ the particle injection
rate (in units of $\mathrm{GeV^{-1} \, cm^{-3} \, s^{-1}}$) and $N'(E')$ the steady particle spectrum (in units of $\mathrm{GeV^{-1} \, cm^{-3}}$).  Here, we do not consider an explicitely time-dependent version of this kinetic equation,
because the low statistics in neutrino observations will not allow for the resolution of  time-dependent features.
For the charged secondaries (pions, muons, kaons) we assume that they lose energy mostly by synchrotron radiation, governed by \equ{synchrotrontime}, and that they escape mostly by decay  
\begin{equation}
t'^{-1}_\mathrm{decay}=\frac{m c^2}{E'\,\tau_0} \, ,
\label{equ:timedecay}
\end{equation}
where $\tau_0$ is the rest frame lifetime.  We show the synchrotron cooling and decay rates for pions, muons, and kaons as a function of energy in \figu{times}, where it can be clearly read off that the synchrotron and decay rates are equal at different energies which depend on the particle species. 
The gray-shaded region shows the estimated range for the escape (or adiabatic cooling) rate if the size of the acceleration region is given by a shell width determined by a variability timescale ranging from $0.01 \, \mathrm{s}$ to $1 \, \mathrm{s}$ (Lorentz factor $\Gamma=300$, redshift $z=2$; \cf, \equ{ism}).
One can see that synchrotron cooling and decay typically dominate,  which is an assumption implied in \Refs~\cite{Waxman:1997ti,Guetta:2003wi}.
As a consequence,  the (steady state) spectrum is loss-steepened by two powers above the energy  $E'_c \propto \sqrt{m^5/\tau_0}$ given by \equ{ec},
where synchrotron cooling and decay rates are equal. Thus, the different
secondaries, which have different masses $m$ and rest frame lifetimes $\tau_0$,  will exhibit different break
energies which solely depend on particle physics properties (and the value of
$B'$). For example, the pion spectrum will  have a break at
\begin{equation}
	E'_{c,\pi} = 1.1 \cdot 10^{6} \, \left( \frac{B^{'}}{10^{5} \, \text{G}} \right)^{-1} \, \giga\electronvolt \quad .
	\label{equ:ecpion}
\end{equation}
Note that in the absence of energy losses $b(E)=0$ . Then we can easily solve the differential equation in \equ{steadstate} in order to obtain 
\begin{equation}
 \label{equ:steadystateesc}
 N'(E')=Q'(E')\,t'_\mathrm{esc} \, .
\end{equation}

\subsection{Typical results}
\label{sec:typ}

Here, we discuss the results qualitatively and quantitatively using the often used Waxman-Bahcall flux~\cite{Waxman:1998yy} as an example; see also \Ref~\cite{Baerwald:2010fk}. 

\subsubsection*{Qualitative shape of the neutrino spectra}
\label{sec:qual}

If the protons are injected with a power law with injection index two, one obtains for the prompt GRB neutrino flux, referred to as ``WB flux'', 
\begin{equation}
	E^2_{\nu} \phi \propto \left\{ \begin{array}{ll} \left( \frac{E_{\nu}}{E_{\nu, \mathrm{break}}} \right)^{\alpha_{\nu}} & \text{for} \; E_{\nu} < E_{\nu, \mathrm{break}} \\ \left( \frac{E_{\nu}}{ E_{\nu, \mathrm{break}} } \right)^{\beta_{\nu}} & \text{for} \; E_{\nu, \mathrm{break}} \leq E_{\nu} < E_{\nu, \pi} \\ \left( \frac{E_{\nu}}{E_{\nu, \mathrm{break}}} \right)^{\beta_{\nu}} \left( \frac{E_{\nu}}{E_{\nu, \pi}} \right)^{-2} & \text{for} \; E_{\nu} \geq E_{\nu, \pi} \end{array} \right.  \label{equ:WB}
\end{equation}
with $\alpha_\nu = \alpha_p -\beta_\gamma + 1 \simeq +1$ and $\beta_\nu = \alpha_p - \alpha_\gamma  + 1 \simeq 0$, where this form applies to the muon neutrinos from pion decays. The values for the break energies used in \Ref~\cite{Waxman:1998yy} are $E_{\nu, \mathrm{break}} \simeq 10^5 \, \giga\electronvolt$ and $E_{\nu, \pi} \simeq 10^7 \, \giga\electronvolt$. For the comparison
to the quantitative computation, it is useful to define a reference parameter set which reproduces this shape.
We use the analytical estimates from \Ref~\cite{Guetta:2003wi}, assuming that $\Gamma = 10^{2.5}$ and $\textit{z} = 2$.  The first break energy $E_{\nu, \mathrm{break}}$ can be related to  $\varepsilon_{\gamma,\text{break}}$ from the  threshold of the photohadronic interactions at the source. As a minor difference to \Ref~\cite{Guetta:2003wi}, where heads-on collisions between photons and protons are assumed for the threshold, we include the effect that the pion production efficiency peaks at higher center-of-mass energies (see Fig.~4 in \Ref~\cite{Hummer:2010vx}) to match the numerical results. This leads to a factor of two higher photon energy break in the SRF ($14.8 \, \mathrm{keV}$). The second break comes from pion cooling in the magnetic field, it can be computed from Eqs.~(\ref{equ:boost}) and~(\ref{equ:ecpion}) taking into account that about 1/4 of the pion energy is deposited per neutrino species. In order to reproduce $E_{\nu, \pi} \simeq 10^7 \, \giga\electronvolt$, one has $B' \simeq 3 \cdot 10^5 \, \mathrm{G}$. For the flux normalization, we use \equ{wb} for this reference parameter set.  We summarize these parameters in \Tab~\ref{tab:params} (last column).

However, as demonstrated in \Ref~\cite{Baerwald:2010fk}, there are qualitative differences
to this analytical picture because: 1. Multi-pion production processes in the photohadronic interactions affect the spectral shape. 2. Magnetic field effects, different for pions, muons, and kaons (\cf, \figu{times}), lead to a characteristic multi-peak spectrum. 3. Flavor mixing and pile-up effects pronounce the peak from muon decays. We will explicitely demonstrate these effects below.

\subsubsection*{A quantitative example}

\begin{figure}[tp]
\begin{center}
\includegraphics[width=\textwidth]{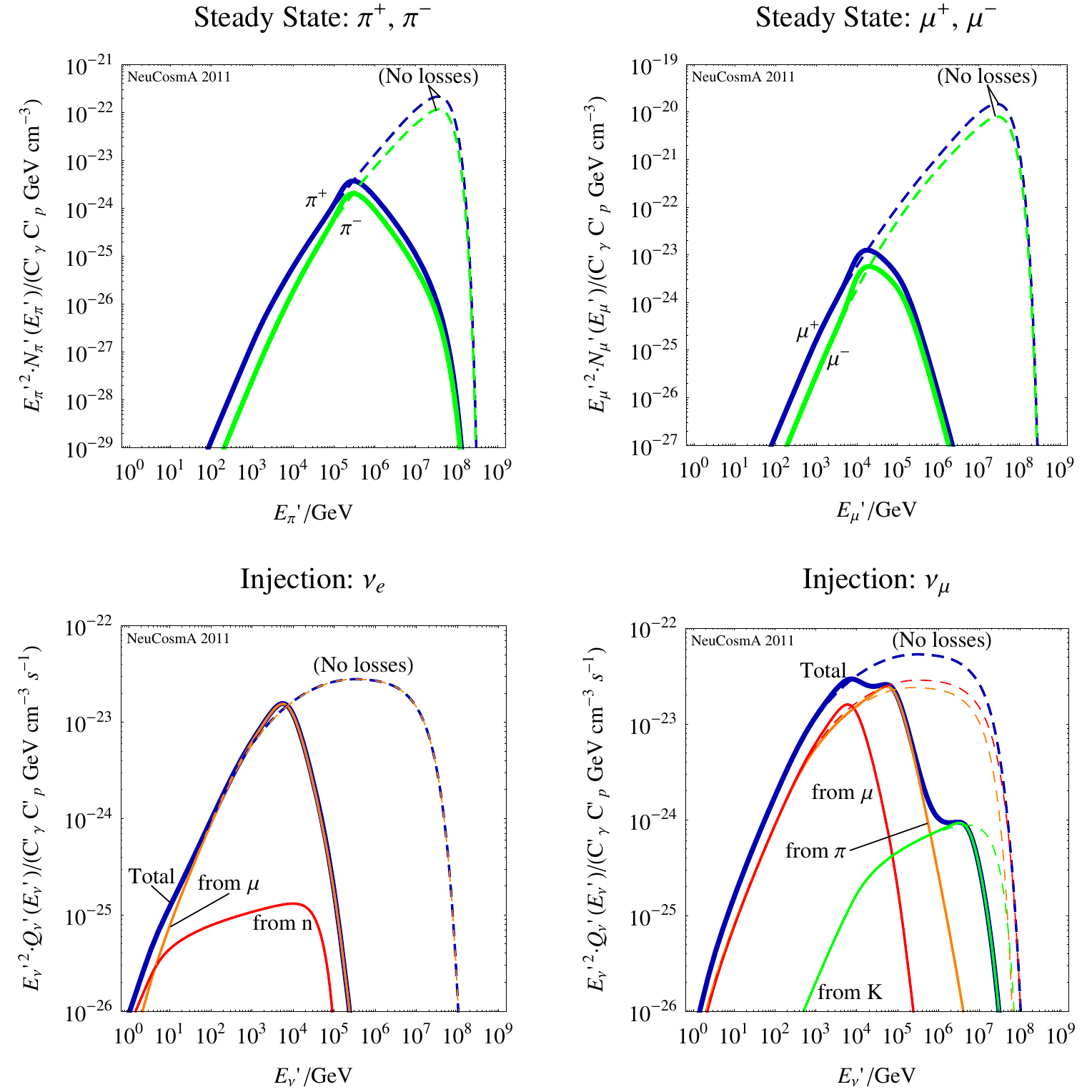} 
\end{center}
\mycaption{\label{fig:example} 
Pion (upper left), muon (upper right), electron neutrino (lower left), and muon neutrino (lower right) spectra in the SRF, as given in the plot labels, for the reference parameters in \Tab~\ref{tab:params}. Solid curves include energy losses of the secondaries, dashed curves are shown without these energy losses.}
\end{figure}

We show the steady pion (upper left) and muon (upper right) spectra, as well as the electron neutrino (lower left), and muon neutrino (lower right) injection spectra for the chosen parameter values (\cf, \Tab~\ref{tab:params}, last column) in the SRF in \figu{example}. The steady pion spectra $N_\pi'(E'_\pi)$ exhibit two breaks: one at about $10^3 \, \mathrm{GeV}$, which comes from the break in the photon spectrum, and one at about $10^5 \, \mathrm{GeV}$, determined by \equ{ecpion}, above which the synchrotron losses dominate. Note that the pion injection and steady state spectra below the second break are connected with  \equ{steadystateesc}, which leads to a different spectral index by one power coming from $t'^{-1}_{\mathrm{decay}}$, see \equ{timedecay}. The steady muon spectra $N_\mu'(E'_\mu)$ are shown in the upper right panel of \figu{example}. Here, there are in fact two cooling breaks visible at about $10^4 \, \mathrm{GeV}$ (muons) and $10^5 \, \mathrm{GeV}$ (pions, from which the muons originate), see \figu{times}. Comparing the solid curves, including energy losses, with the dashed curves, without energy losses,  the pile-up effect (the part of the spectrum where the flux including cooling is actually higher than without cooling) is more pronounced for the muons than for the pions.

In the lower two panels of \figu{example}, we show the electron (left) and muon (right) neutrino injection spectra at the source, where the contributions from the individual parents are marked (\cf, \figu{flowchart}). Since electron neutrinos or antineutrinos (left panel) can only be produced in muon or neutron decays, the muon decays clearly lead to a pronounced peak. The neutron decays only contribute significantly at very low energies. The muon neutrinos (right panel) can be produced from muon, pion, or kaon decays. One can easily see in this figure that the different critical energies in \equ{ec} (see also \figu{times}) lead to a characteristic triple-peak structure, where the kaons are least affected by the synchrotron cooling. 

\begin{figure}[t!]
\begin{center}
\includegraphics[width=\textwidth]{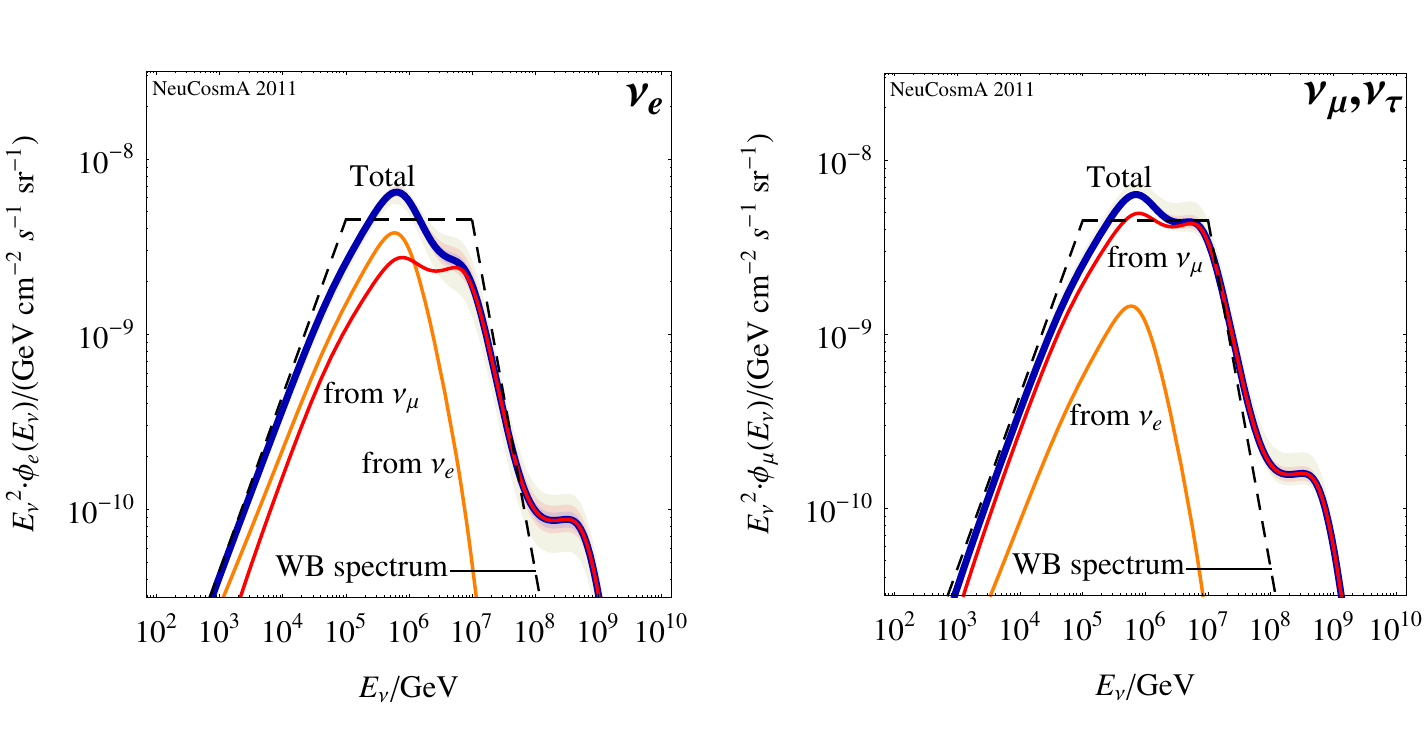}  
\end{center}
\mycaption{\label{fig:fluxobs} Electron (left panel) and muon/tau (right panel) neutrino fluxes at the detector including Lorentz boost, redshift, and flavor mixing for the reference parameters in \Tab~\ref{tab:params} (thick solid curves). The individual contributions from the spectra before flavor mixing are shown as well (thin solid curves). The neutrino energy flux density of the total fluxes (thick curves) is normalized to Eqs.~(\ref{equ:wb}) and~(\ref{equ:WB}), shown as ``WB spectrum''.
The shaded regions show the impact of the $3 \sigma$ mixing angle uncertainties now, in about 2015, and in about 2025 (see main text for details). The uncertainty in 2025 is basically smaller than the thickness of the curve.}
\end{figure}

In \figu{fluxobs}, we include the effects of flavor mixing, Lorentz boost, and redshift. Here, the electron (left panel) and muon or tau (right panel) neutrino fluxes at the detector (excluding Earth attenuation) for the reference parameters in \Tab~\ref{tab:params} are shown (thick solid curves). The individual contributions from the spectra before flavor mixing are shown as well (thin solid curves), which correspond to the thick solid curves in the lower panels of \figu{example}, correctly weighted.  The neutrino energy flux density of the total fluxes (thick curves) is normalized to Eqs.~(\ref{equ:wb}) and~(\ref{equ:WB}), shown as ``WB spectrum''. For the electron neutrino spectrum, a clear dominant peak can be seen, since the main contribution comes from the $\nu_e$ spectrum before flavor mixing. For the muon/tau neutrino spectra, which are representative for muon tracks in neutrino telescopes, flavor mixing pronounces the peak from muon decays somewhat. However, the peak flux is dominated by a double peak at around the WB plateau, and the original shape is roughly reproduced. The additional kaon decay hump leads to a triple-peak spectrum. Note that even the neutrino spectrum from pion decays is not flat at the WB plateau, since high-energy photohadronic processes change the spectral shape, whereas the original shape can be exactly reproduced if these processes are switched off~\cite{Baerwald:2010fk}.

Since in \equ{flmix} the neutrino mixing angles appear, the uncertainties of the mixing angles will lead to uncertainties in the fluxes and flavor ratios. In order to discuss these, we use the following values and current $3 \sigma$ ranges (see \Ref~\cite{Schwetz:2008er}): $\sin^2 \theta_{23}=0.5$ ($3\sigma$: 0.36 ... 0.67), $\sin^2 \theta_{12}=0.318$ ($3 \sigma$: 0.27 ... 0.38), $\sin^2 \theta_{13}=0$ ($3\sigma$: $\sin^2 \theta_{13} \le 0.053$). In addition, we consider future improved bounds ($3\sigma$) on $\theta_{13}$ from Daya Bay $\sin^2 \theta_{13} \le 0.012$ and $\theta_{23}$ from T2K $0.426 \le \sin^2 \theta_{23} \le 0.574$, which represent the expected sensitivities at around 2015~\cite{Huber:2009cw} if $\theta_{13} \simeq 0$, which we assume for the sake of simplicity. Beyond these, a neutrino factory may improve the bounds even further: $\sin^2 \theta_{13} \lesssim 1.5 \cdot 10^{-5}$~\cite{Huber:2003ak} and $0.46 \le \sin^2 \theta_{23} \le 0.54$~\cite{Tang:2009na} ($3\sigma$), which we label ``2025''. For $\theta_{12}$, the errors may improve somewhat, but new generations of experiments are not assumed here. Since $\theta_{12}$ is of secondary importance in \equ{flmix}, we do not consider such future improvements. The shaded regions in \figu{fluxobs} show the ranges for the flux including these $3 \sigma$ uncertainties, where the different shadings correspond to the current errors, the ones in ``2015'', and the ones in ``2025''. Obviously, there is currently some uncertainty in the fluxes in the pion- and kaon-decay dominated ranges, where the $\nu_\mu$ production dominates, especially for the $\nu_e$ flux at the observer. However, already the improved knowledge from the upcoming experiments will drastically reduce this uncertainty. If information from a neutrino factory is available, the errors collapse to within  the thick curves.

\begin{figure}[t!]
\begin{center}
\includegraphics[width=0.5\textwidth]{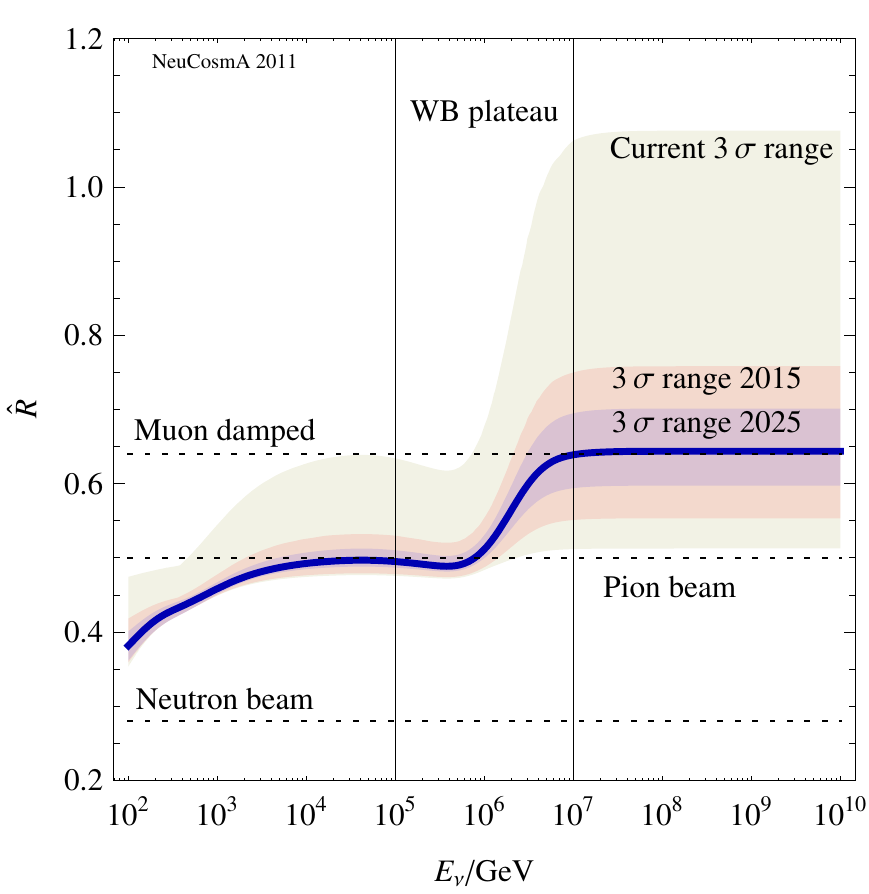} 
\end{center}
\mycaption{\label{fig:flrobs} Flavor ratio $\rhat$ as a function of $E_\nu$ at the detector.  Here, ``Pion beam'' refers to a flavor ratio $\phi_e:\phi_\mu:\phi_\tau=1:2:0$ before flavor mixing, ``Muon damped'' to $0:1:0$, and ``Neutron beam'' to $1:0:0$. 
The dashed regions show the impact of the $3 \sigma$ mixing angle uncertainties now (``current''), in about 2015, and in about 2025 (see main text for details).}
\end{figure}

Finally, we show in \figu{flrobs} the flavor ratio $\rhat$ as a function of $E_\nu$ at the detector. One can easily see the flavor transition from ``pion beam'', where  the whole decay chain in \equ{piplusdec} or \equ{piminusdec} is present, to a ``muon damped source'', where the muons lose energy faster than they decay and the muon decay part is not present.
In the presence of flavor and magnetic field effects, our results are consistent with \Ref~\cite{Lipari:2007su}, in spite of the slightly different methods; see also \Ref~\cite{Hummer:2010vx}. Again, we show the uncertainty coming from the mixing angles as dashed regions. Here, on a linear scale, the effect is somewhat larger than for the flux. However, again, in about 2015 the different regimes (pion beam and muon damped) can, in principle, be discriminated by the value of $\rhat$.

In \Sec~\ref{sec:diffuse}, we use \figu{fluxobs}, right panel, and \figu{flrobs} as the representatives for the observables.

\subsection{Limitations of the model, comparison to other approaches}
\label{sec:lim}

Our model can be regarded as minimal approach reproducing the results from \Refs~\cite{Waxman:1997ti,Guetta:2003wi,Abbasi:2009ig}, which means that it also shares some of the weaknesses with these approaches. First of all, in all these approaches the proton spectrum and target photon field are put in ``by hand'', which is motivated from observations, but may not be self-consistent for the individual burst. For example, the photohadronic cooling is not taken into account, compared to more detailed numerical simulations as in \Ref~\cite{Murase:2005hy}. In the target photon spectrum, additional radiation processes may be at work, such as proton synchrotron radiation, which are not described by our approach (see, \eg, \Refs~\cite{Becker:2010cj,Gao:2011jt}). In addition, the effects of adiabatic cooling and additional interactions are neglected for the secondaries, which may change the spectral shape in extreme cases. 

Compared  to \Refs~\cite{Waxman:1997ti,Guetta:2003wi,Abbasi:2009ig}, our model treats the energy losses of the secondaries explicitely, which is important in the presence of strong magnetic fields to describe the spectrum and flavor ratios accurately, including pile-up effects which cannot be modeled in analytical treatments. Whereas additional processes affecting the primaries (protons, photons) may change the spectral shape, it is important to note that the flavor ratio relies on the description of the secondaries, which mostly depends on (known) particle physics. In addition, we include additional pion production modes in the photohadronics, \ie, direct ($t$-channel) production, higher resonances, and multi-pion production, and the helicity-dependent muon decays, which have some impact on the flavor ratio. As it is explicitely demonstrated in \Refs~\cite{Hummer:2010vx,Baerwald:2010fk}, it is fair to say that the $\Delta(1232)$ resonance (in the particle physics sense) hardly ever dominates the charged pion production in GRBs, see Fig.~5 in \Ref~\cite{Hummer:2010vx}. Although the methods are different, our ingredients are similar to \Ref~\cite{Lipari:2007su}, which we were able to reproduce and confirm, but perhaps far more efficient.

\section{Variation of luminosity and impact of gamma-ray detection threshold}
\label{sec:threshold}

One may argue that the threshold for the $\gamma$-ray detection may lead to a selection bias in a stacking analysis. This is not taken into account in \Secs~\ref{sec:redshift} and~\ref{sec:lowstatistics}, which means that the low luminosity GRBs at high redshifts may be over-counted there. This threshold is (for {\em Swift}) shown as gray-shaded area in Fig.~1 of Kistler et al.~\cite{Kistler:2009mv} and roughly excludes the region $L_{\mathrm{iso}}/(4 \pi d_L^2(z)) \lesssim const$ in the $L_{\mathrm{iso}}$-$z$-plane. Note that this detection threshold may potentially lead to a qualitative difference between the diffuse and quasi-diffuse (from the stacking using the gamma-ray counterpart) neutrino fluxes.

We have tested the impact of this threshold for several assumptions of the luminosity distribution function by simultaneously generating redshift and luminosity distributions. First of all, it should be noted that we do not observe a significant impact of the threshold on the contribution function or neutrino spectra as long as the stacked bursts satisfy $L_{\mathrm{iso}} \gtrsim 10^{51} \, \mathrm{erg} \, \mathrm{s}^{-1}$, \ie, a cut on the luminosity is applied. Since this cut is exactly the one which has been also used in \Ref~\cite{Kistler:2009mv} to derive the correction to the star formation rate evolution of the {\em Swift} GRBs, our redshift distribution is based on, this result is not surprising. Note that, of course, the cut on the luminosity depends on the satellite used for the stacking sample. As a qualitative difference, if an additional luminosity distribution is applied, the cumulative distribution function of the bursts  will be broader (the statistical errors add, to a first approximation, in a Gaussian way), which means that the systematical errors given in \Sec~\ref{sec:lowstatistics} on the stacked flux can be regarded as a lower limit. The details, however, depend on the assumed  luminosity distribution, see below for an example, which is why we do not include the luminosity in the main text. 

\begin{figure}[t]
	\centering
	\includegraphics[width=\textwidth]{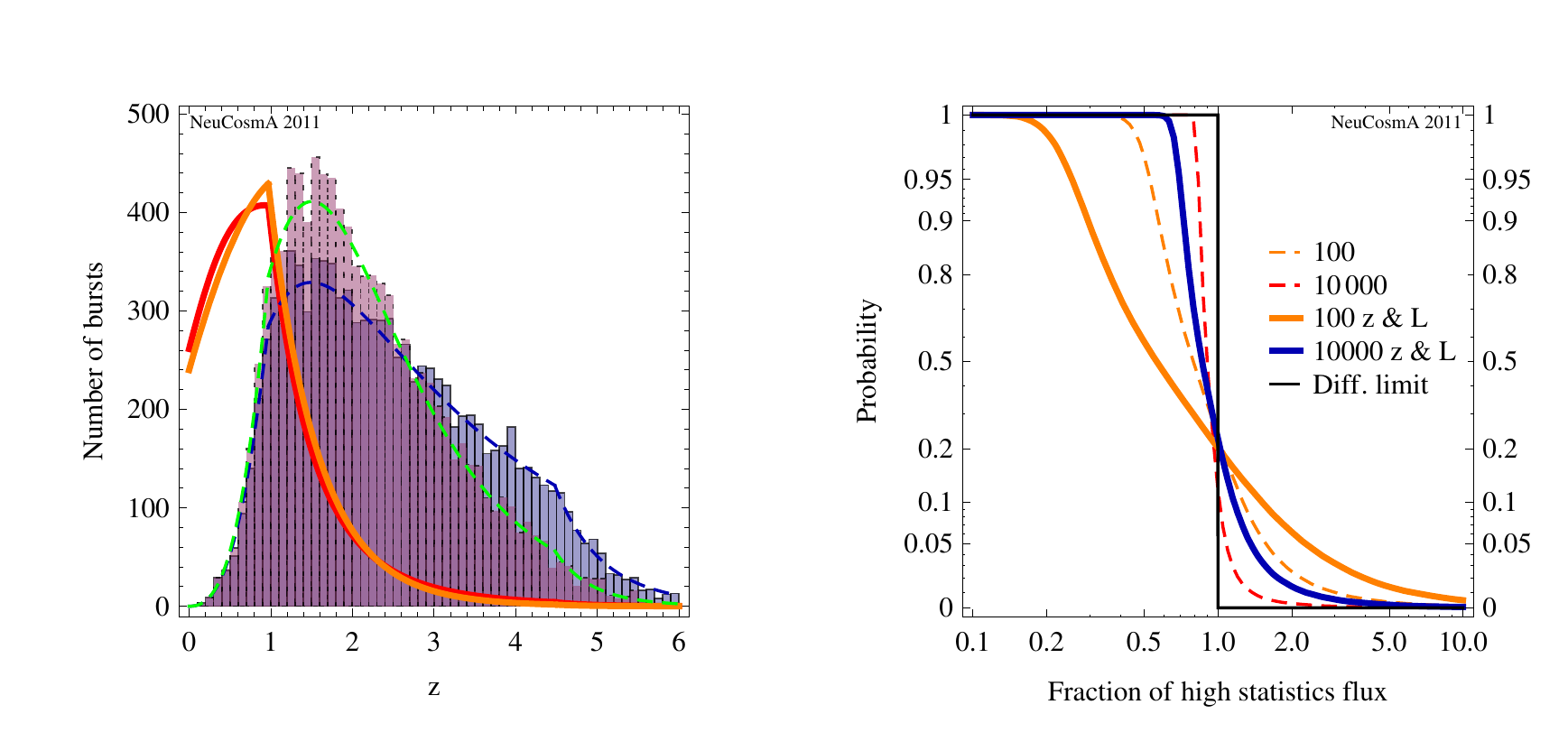}
	\mycaption{\label{fig:zLvariation1}
In the left panel, we show the distribution of the number of observable GRBs as a function of redshift (histograms) with the luminosity being integrated out, where the gamma-ray detection threshold is taken into account. The two shown $10\,000$ burst samples differ in their distribution in luminosity, with the dark one being wider ($\sigma_{x_1} = 1.0$) than the light one being narrower ($\sigma_{x_2} = 0.5$), see main text. The dashed curves depict the analytical result of the distribution in redshift, the thick solid curves the corresponding contribution functions (red/dark: wide, orange/light: narrow). Notice that the contribution functions are in a.u. and have been normalized to 1 when integrated over the depicted redshift range. In the right panel, the cumulative distribution function similar to \figu{lowstatz} is shown for $n=100$ and $n=10 \, 000$ for only redshift varied (dashed) and redshift and luminosity varied simultaneously (solid), where the wide luminosity distribution is used.
}
\end{figure}

If bursts with $L_{\mathrm{iso}} \lesssim  10^{51} \, \mathrm{erg} \, \mathrm{s}^{-1}$ are included in the stacking analysis, \ie, bursts which may be visible for small $z$ but not for high $z$, the luminosity distribution may have an impact.
Assume, for instance, that the GRB luminosities follow a Gaussian distribution in the exponent $L_{\mathrm{iso}} = 10^{x} \, \text{erg} \, \second^{-1}$ and a central value $\overline{x} \simeq 51$, see Fig.~1 in \Ref~\cite{Kistler:2009mv}. Consider a wider standard deviation $\sigma_{x_1} = 1.0$, and a narrower one $\sigma_{x_2} = 0.5$ for comparison.\footnote{There have been some studies to extract the luminosity distribution of the {\em Swift} sample, see \Refs~\cite{Wanderman:2009es,Cao:2011ar}. For instance, the broken power law distribution in \Ref~\cite{Wanderman:2009es} leads to results between our two extreme cases.}
Both of these functions are expected to have a significant effect on the result of the observed number of bursts compared to \figu{redshiftdistr} if the {\em Swift} detection threshold is applied. We show the results in \figu{zLvariation1}. In the left plot, the redshift distributions with luminosity integrated out are shown, together with the corresponding contribution functions.
As one can see from the plot,  the high redshift contribution is suppressed for the narrower distribution, because not as many bursts exceeding the threshold as for the wide distribution  are generated in that region. However, the luminosity-integrated contribution functions both peak at $z \simeq 1$ and are very similar to the one in \figu{redshiftdistr}. As far as the cumulative distribution is concerned (right panel of \figu{zLvariation1} for the wide distribution function of luminosity), the spreads including the luminosity distribution (solid curves) are significantly larger than for the redshift distribution (dashed curves) only, which is expected from statistics. In fact, in this example, the 90\%CL error on the stacking of 100 bursts is roughly 100\%. That number, however, depends on the breadth of the luminosity distribution of the stacked bursts, which is chosen aggressively wide in this example. In either case, the redshift distribution will limit this systematical error, as we argue in the main text.

\end{document}